\documentclass[preprint,12pt]{elsarticle}

\usepackage{amsmath,amsfonts,amssymb,amsthm,dsfont}
\usepackage{array}
\usepackage{caption}
\usepackage{subcaption}
\usepackage{hyperref}
\usepackage{lineno}

\newcommand{\Av}[1] {\underset {#1} {\rm Ave}\,}
\newcommand{\E} {{\mathbb E}}
\newcommand{\Z} {{\mathbb Z}}
\newcommand{\R} {{\mathbb R}}
\newcommand{\C} {{\mathbb C}}

\newcommand{\om} {{\omega}}
\newcommand{\Cov} {{\rm Cov}}

\newtheorem{theorem}{Theorem}

\newtheorem{proposition}[theorem]{Proposition}

\journal{\textbackslash}

\begin{document}

\begin{frontmatter}



\title{Scale Dependencies and Self-Similar Models with Wavelet Scattering Spectra}


\author[ens]{Rudy Morel\corref{corr}}
\ead{rudy.morel@ens.fr}
\cortext[corr]{Corresponding author.}
\author[ens]{Gaspar Rochette}
\author[ens]{Roberto Leonarduzzi}
\author[cfm]{Jean-Philippe Bouchaud}
\author[CdF,ens]{Stéphane Mallat}

\affiliation[ens]{
organization={École Normale Supérieure},
addressline={45 rue d'Ulm},
postcode={75005},
city={Paris},
country={France}}
\affiliation[CdF]{
organization={Collège de France},
addressline={11 place Marcelin Berthelot},
postcode={75231},
city={Paris},
country={France}}
\affiliation[cfm]{
organization={Capital Fund Management},
addressline={23 Rue de l'Université},
postcode={75007},
city={Paris},
country={France}}

\begin{abstract}
We introduce the wavelet scattering spectra which provide non-Gaussian models of time-series having stationary increments. A complex wavelet transform computes signal variations at each scale. Dependencies across scales are captured by the joint correlation across time and scales of wavelet coefficients and their modulus. This correlation matrix is nearly diagonalized by a second wavelet transform, which defines the scattering spectra. We show that this vector of moments characterizes a wide range of non-Gaussian properties of multi-scale processes. We prove that self-similar processes have scattering spectra which are scale invariant. This property can be tested statistically on a single realization and defines a class of wide-sense self-similar processes. We build maximum entropy models conditioned by scattering spectra coefficients, and generate new time-series with a microcanonical sampling algorithm. Applications are shown for highly non-Gaussian financial and turbulence time-series
\footnote{This work is supported by the PRAIRIE 3IA Institute of the French ANR-19-P3IA-0001 program and the ENS-CFM models and data science chair.}
\footnote{Declarations of interest: none.}
.
\end{abstract}



\begin{keyword}
maximum entropy \sep 
non-Gaussianity \sep
scattering \sep
self-similarity \sep
time-series \sep
wavelets.


\end{keyword}

\end{frontmatter}


\section{Introduction}
Time-series having stationary increments with variations on a wide range of scales are encountered in physics, finance, biology, medicine and many other fields.
Such multi-scale time-series  typically include complex intermittent phenomena with local bursts of activity, and time-asymmetries due to some form of causality.
The importance of this topic was first recognized by Mandelbrot 
\cite{mandelbrot1968fractional,mandelbrotmultifractals,mandelbrot1982fractal} and led to a considerable body of work on multifractal signals
\cite{bacry1993singularity,muzy1994multifractal,abry2000wavelets,jaffard2004wavelet,jaffard2006wavelet,wendt2009wavelet,leonarduzzi2018generalized}. 
Among multi-scale processes, self-similar models
have a probability distribution which is invariant to scaling, up to multiplicative factors. 
To validate numerically
such models, it is however
necessary to introduce weaker forms of self-similarity that can be estimated over limited data.

Simplified multi-scale models have been introduced from
marginal distributions of signal increments,
by Frisch and Parisi \cite{FrischParisi80}.
They define a weak form of self-similarity 
from
a scale invariance of high order moments of these marginal distributions. This
can be sufficient to detect non-Gaussian distributions.
Section \ref{SecMarginalMoments} reviews these models together
with the multifractal formalism, which replaces
increments by wavelet coefficients. 
Marginal distributions at each scale are simple to estimate, but they
do not capture dependencies of signal variations across scales.
These dependencies are crucial to specify many properties, in particular,
the existence of transient events, which have particular signatures at multiple scales. 

Models of multi-scale distributions can be defined as a maximum entropy distribution 
conditioned by a vector of moments $\E\{\Phi(X)\}$. 
If they exist, they have an exponential probability distribution
\begin{equation}
\nonumber
\label{exponmod}
p_\theta (x) = Z^{-1}_\theta \, e^{-\theta^T\, \Phi(x)}.
\end{equation}
for $\theta\in\R^M$, where $M$ is the number of moments. 
Maximum entropy models depend only on the energy vector $\Phi(x)$, which needs to be chosen appropriately. Gaussian processes are maximum entropy models conditioned by first and second order moments.

A central result of the paper is the construction of $\Phi$, so that 
$\E\{\Phi(X)\}$ specifies scale dependencies, and provide accurate models of multi-scale time-series. The dimension $M$ of $\Phi(X)$ is much smaller than the
dimension of $X$, so that it can define a consistent mean estimator which
converges to $\E\{\Phi(X)\}$ when the
dimension of $X$ increases. As a result, maximum entropy models can be estimated from 
a single realization of $X$. New samples of $X$ are generated by sampling a microcanonical
model, which approximates the macrocanonical model \cite{bruna2019multiscale}.

A wavelet transform computes multi-scale signal variations. 
Complex wavelet coefficients carry a complex modulus and a complex phase information.
Section \ref{SecDependenciesAcrossScales} proves that wavelet coefficients
are nearly uncorrelated at different scales, because their phases oscillate at different frequencies. To measure non-linear dependencies across scales, it is tempting to move towards higher order moments \cite{brillinger1965introduction}.
This requires to compute many moments with high variance estimators, which gives poor numerical results over limited size time-series.
Lower variance estimators have been studied by replacing high order moments with phase harmonics \cite{mallat2020phase} or by eliminating the phase with a modulus non-linearity \cite{bruna2013invariant, portilla2000parametric}. We show that
scale dependencies can be 
captured
by correlating wavelet coefficients and their modulus.
We prove that self-similar processes yield normalized correlation matrices
which are invariant to scaling.
Section \ref{SubSecWideSenseSelfSimilarity} derives a definition
of wide-sense self-similarity, which
is analogous to the definition of wide-sense stationarity, where invariance to translation of correlation matrices is replaced by an invariance to scaling.

Wavelet modulus cross-correlation matrices are too large to be estimated 
accurately from a single time-series realization. 
Section \ref{SubSecScatteringCovariance} shows that applying a second wavelet transform defines
a scattering covariance matrix which is nearly diagonal. 
Dependencies across scales are captured by diagonal
scattering cross-correlation coefficients, 
also called scattering cross-spectrum, 
which can be
estimated from a single realization.
We shall see that the scattering spectra provide an
interpretable dashboard which captures non-Gaussian properties, including bursts of activity and time-asymmetries, as well as self-similarity.

Fractional Brownian motions, Poisson processes, multifractal random walks and Hawkes processes
are often used as models of multi-scale processes which may or may not be self-similar.
Section \ref{SecNumericalDashboard} shows that the scattering spectra
reveal their specific properties. By analyzing the scattering spectra of
S\&P financial time-series and turbulent jets, we show that none of the mathematical models presented captures all properties of these complex time-series.

Section \ref{SecSynthesisByEntropyMax} defines 
maximum entropy models conditioned by scattering spectra values.
We generate time-series according to these models with the microcanonical sampling
algorithm in \cite{bruna2019multiscale}.
We show that these generative models can approximate 
fractional Brownian motions,  multifractal random walks and Hawkes processes but also 
S\&P financial time-series or turbulent jets. 
The code used in numerical experiments is available at \url{https://github.com/RudyMorel/scattering_spectra}.

{\bf Notations:} We write $M^T$ the adjoint and hence the complex conjugate of the transposed matrix $M$. We write $\widehat{x}(\om)$ the Fourier transform of $x(t)$.

\section{Multi-scale Moments}
\label{SecMarginalMoments}

We consider a multi-scale random process $X(t)$ whose increments are stationary. 
Gaussian models are maximum entropy models conditioned by first and second order moments. In order to capture non-Gaussian and self-similar properties, one can compute
 higher order moments of increments.
The section reviews the scaling properties of these moments.

\subsection{Self-Similarity and Power Spectrum}
\label{subsec:power-spectrum}

If $X(t)$ is stationary then its increments are stationary but the reverse is not always true. 
For example, a Brownian motion $X$ 
has stationary increments but $\E\{X(t)\}$ and $\E\{X^2(t)\}$ depend on $t$ \cite{Taqqu}.
The increments of a random process $X(t)$ at intervals
$2^{j} \in \R^+$ for $j \in \R$ are written
\[
\delta_j X(t) = X(t) - X(t-2^{j}) .
\]
The lag $2^j$ can also be interpreted as a scale parameter. 
We suppose that $\delta_j X$ is stationary for any $j \in \R$. 

Mandelbrot et al. \cite{mandelbrot1997multifractal} introduced a strong definition of self-similarity
from the joint distribution of increments. A process $X$ is said to be self-similar \cite{mandelbrot1997multifractal}
up to a maximum scale $2^J$ if for all $\ell \geq 0$ 
there exist real random variables $A_\ell$ which are log infinitely divisible and independent of $X$ such that
\begin{equation}
	\label{strongSelfSimilaritydX}
	\Big\{ \delta_j X(t)  \Big\}_{j\leq J, t \leq N}
	\overset{d}{=} 
	A_\ell \ \Big\{ \, \delta_{j-\ell}X(2^{-\ell}t) \Big\}_{j\leq J, t \leq N} .
\end{equation}
This equality is in distribution, which
means that joint probability distributions of random variables on the left and right hand-sides are equal
for any $(j_1,...,j_n)$ and $(t_1,...,t_n)$ with $n > 0$.
Increments thus have joint distributions which are invariant to dilation, up to random multiplicative factors. The maximum scale $2^J$ is called the {\it integrable scale}. It may be fixed, in that case
we assume that $X(t)$ and $X(t-\tau)$ are nearly independent for $\tau \gg 2^J$.

If increments are stationary then 
their auto-correlation $\E\{\delta_j X(t)\, \delta_{j'} X(t-\tau)\}$ 
only depends on $\tau$. By renormalizing its Fourier transform along $\tau$, 
one can mathematically define  a generalized power spectrum $P_X(\om)$ of $X$ \cite{Taqqu}. 
The non-stationarity of $X$ appears as a singularity of $P_X(\om)$, which tends to $\infty$ 
at $\om = 0$. This power spectrum specifies second order moments of increments. 

With a scaling argument, one can prove \cite{Taqqu} that self-similar processes 
have a power spectrum which is also self-similar and thus has a power-law scaling
\begin{equation}
\label{eq:pl-ps}
P_X (\om) = c_2\, |\om|^{-\zeta_2-1}
\end{equation}
which is singular at $\om = 0$.

\subsection{Increment High Order Moments}
\label{SubSecHighOrderMomentsandSelfSimilarity}

First and second order moments define Gaussian maximum entropy models. 
In order to build non-Gaussian multi-scale models, one can compute
$q$ order moments of increments, if they exist:
\begin{equation}
	\label{variationsOrderq}
	\forall j\in\R ~~ , ~~ \E\{|\delta_j X(t)|^q\}.
\end{equation}
For self-similar processes, these multi-scale moments have a power-law scaling
\begin{equation}
	\label{marginalSelfSimilaritydX}
	\E\{|\delta_j X(t)|^q\} = \widetilde c_q\, 2^{j \zeta_q} .
\end{equation}
If $X$ is Gaussian and self-similar then one can verify that
$\zeta_q$ is linear in $q$ \cite{Taqqu}.
It results that any non-linear dependency of $\zeta_q$ as a function of $q$
implies that $X$ is not Gaussian. This was initially proposed by Kolmogorov as a test to detect
non-Gaussian properties in turbulent flows.
Under
appropriate hypotheses, the multifractal theory \cite{jaffard2004wavelet} proves that (\ref{variationsOrderq})
 specifies the pointwise Holder regularity of X, through a
spectrum of singularity. 

The moment power-law scaling (\ref{marginalSelfSimilaritydX}) is a weak form of self-similarity
which 
can be tested statistically.
On the other hand, the strong
self-similarity definition (\ref{strongSelfSimilaritydX}) is highly restrictive, often not satisfied, and impossible to
be tested on a single realization.
\ref{AppendixDistributionImpliesMarginal} shows that the strong distribution self-similarity (\ref{strongSelfSimilaritydX}) implies the weak moment self-similarity (\ref{marginalSelfSimilaritydX}). This scaling is simple to test numerically but
is a relatively weak characterization of self-similarity.
The high order increment moments (\ref{marginalSelfSimilaritydX})
remain unchanged when computed on $X(-t)$, and hence do not detect time-asymmetries.
They do not either capture dependencies of increments at different scales $2^j$.
Section \ref{SecDependenciesAcrossScales} introduces a stronger wide-sense definition of self-similarity, which relies on multi-scale moments that depend upon joint time-scale dependencies of $X$.

\subsection{Estimation and Wavelet Transform}
\label{SubSecWavelets}

Defining consistent estimators of moments with fast convergence is necessary to compute
maximum entropy models from a single realization of $X$. 
It has been proved that
a wavelet transform yields nearly optimal estimators of second order moments for self-similar
processes \cite{flandrin1992,wornell1993,masry1993,mccoy1996wavelet}. We thus replace increments by a wavelet transform, whose properties are briefly reviewed.

\subsubsection{Wavelet transform}
A wavelet $\psi(t)$ has a fast decay away from $t=0$, polynomial or exponential for example,
and a zero-average $\int \psi(t)\,{\rm d}t = 0$.
We normalize $\int |\psi(t)|^2 \,dt = 1$.
The wavelet transform computes the variations of a signal $x$
at each scale $2^j$ with
\begin{equation}
\nonumber
W x(t,j) = x\star\psi_j(t)~~\mbox{where}~~\psi_j(t) = 2^{-j}\psi(2^{-j}t).
\end{equation}
If $\psi = \delta(t) - \delta(t-1)$ then it computes signal increments
$W X (t,j) = \delta_j X(t)$. 
To relate regularity properties of signals from 
their wavelet coefficients, it is necessary to use wavelets
having a better frequency localization than a difference of Diracs \cite{jaffard2004wavelet}.
We use a complex wavelet $\psi$ having
a Fourier transform 
$\widehat{\psi}(\om) = \int \psi(t)\, e^{-i \om t}\, {\rm d}t$ which is real, and whose
energy is mostly concentrated at frequencies $\om \in [{\pi}, {2}\pi]$. 
It results that $\widehat{\psi}_j(\omega) = \widehat{\psi}(2^j\omega)$
is non-negligible mostly in $\om \in [2^{-j} \pi, 2^{-j+1}\pi]$.
We suppose that $\psi$
has $m \geq 1$ vanishing moments, which means 
that $|\widehat{\psi}(\om)| = O(|\om|^{m})$ in the neighborhood of $\om = 0$. We will refer to the modulus and complex phase of $Wx(t,j)$ as the \textit{amplitude} and \textit{phase} of the complex wavelet coefficient. 

In the following we shall restrict the scales $2^j$ to dyadic scales, and hence 
$j$ to integers. 
The wavelet transform $W$ satisfies an energy conservation law \cite{mallat1999waveletbook}
specified in \ref{AppendixWaveletProperties}, which implies that it is invertible.

\begin{figure}
\centering
\includegraphics[width=0.6\linewidth]{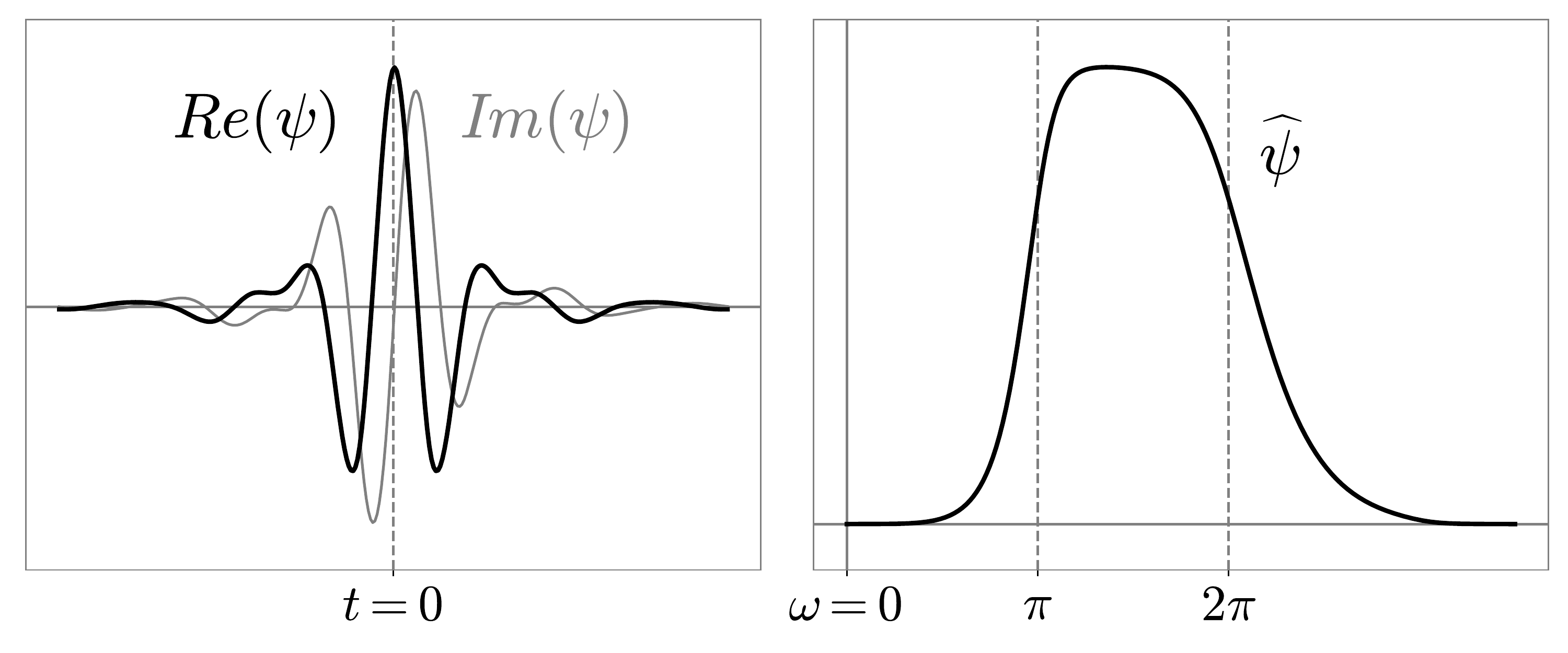}
\caption{Left: complex Battle-Lemarié wavelet $\psi(t)$ as a function of $t$.
Right: Fourier transform $\widehat{\psi}(\omega)$ as a function of $\om$.}
\label{FigMotherWavelet}
\end{figure}

All numerical calculations below are performed with a complex Battle-Lemarié wavelet \cite{battle1987block, lemarie1988ondelettes}, restricted to positive frequencies. Figure \ref{FigMotherWavelet} shows the real and imaginary parts of $\psi$ 
as well as its Fourier transform.
It has an exponential decay away from $t=0$, it has $m = 4$ vanishing moments and satisfies
an energy conservation law
(\ref{littlewood}). If the input signal is sampled at $t \in \Z$ then 
we can only compute wavelet coefficients for $2^j > 1$ and hence $j \geq 1$. 

\subsubsection{Wavelet covariance and spectrum}
Since $\int \psi_j(t) dt = 0$ it results that
$\E\{X \star \psi_j(t)\} = 0$. If $X$ has stationary increments then one can show \cite{Taqqu}
that wavelet coefficients are jointly stationary. 
Their covariance across time and scale
can be written from the power spectrum of $X$:
\begin{equation}
\label{eq:wx-correlation}
\E\{X \star \psi_j (t)  \, X \star \psi_{j-a} (t- 2^j \tau)^*\} = \frac 1 {2 \pi}
\int P_X (\om)\,\widehat{\psi}(2^j \om)\, \widehat{\psi}^*(2^{j-a} \om)\, e^{i \tau 2^j \om}\, d\om ,
\end{equation}
for time-lag $2^j\tau\in\R$ and scale-lag $a\in\Z$.
This covariance becomes negligible when $|a| > 0$ for which the supports of $\widehat{\psi}(\om)$ and $\widehat{\psi}(2^a \om)$ barely overlap.
Indeed, the phases of
$X \star \psi_j$ and $X \star \psi_{j-a}$ vary at different rates, which cancels their correlation. 
For processes $X$ with a power-law decaying power-spectrum (\ref{eq:pl-ps}),
one can prove that such covariance has a polynomial decay away from $\tau=0$ and an exponential decay away from $a=0$ \cite{wornell1993}.
As shown on figure \ref{FigCW}, these coefficients are negligible for distant scales $|a|>1$ and have small non-zero values for $|a| = 1$ because wavelets have a
small frequency overlap.

The diagonal covariance values define a {\it wavelet spectrum}:
\begin{equation}
\label{wavepsecnfsd}
\sigma_W^2(j) = \E\{|X \star \psi_j(t)|^2\} =  \frac 1 {2 \pi}
    \int P_X (\om)\,|\widehat{\psi}(2^j \om)|^2\, d\om .
\end{equation}
It integrates $P_X(\om)$ over the
frequency intervals $[2^{-j} \pi , 2^{-j+1} \pi]$, where $\widehat \psi(2^j \om)$ is mostly supported.
It does not depend upon $t$ because of stationarity, and is thus estimated by an empirical average
\begin{equation}
\label{waveletspectestima}
\widetilde \sigma_W^2 = \Av{t}(|X \star \psi_j(t)|^2).
\end{equation}

\subsubsection{Self-similar wavelet coefficients}
If $X$ is strongly self-similar according to (\ref{strongSelfSimilaritydX}) then \ref{AppendixDistributionImpliesMarginal} derives that 
\begin{equation}
	\label{strongSelfSimilarityWX}
	\forall \ell\geq 0 ~,~ \Big\{ X\star\psi_{j}(t) \Big\}_{j\leq J, t \leq N} 
	\overset{d}{=} 
	A_\ell \ \Big\{ X\star\psi_{j-\ell}(2^{-\ell}t) \Big\}_{j\leq J, t \leq N}.
\end{equation}
\ref{AppendixDistributionImpliesMarginal} also proves 
that wavelet moments of self-similar processes have the same scaling properties
as increments in (\ref{marginalSelfSimilaritydX}). 
For all $q$ such that the moments are defined, there exists
$c_q$ such that
\begin{equation}
	\label{marginalSelfSimilarityWX}
	\forall j, \quad \E\{|X\star\psi_j|^q\} = c_q \,2^{j\zeta_q}.
\end{equation}

\section{Dependencies Across Scales with Phase-Modulus Wavelet Correlations}
\label{SecDependenciesAcrossScales}

We saw that wavelet coefficients of stationary processes are nearly uncorrelated across scales.
Yet, next section shows that non-Gaussian processes have strong dependencies across scales.
Section \ref{SubSecJointWaveletModulus} captures these dependencies by correlating complex wavelet coefficients and their modulus.
Section \ref{SubSecWideSenseSelfSimilarity} shows that
self-similar processes have 
a  normalized phase-modulus wavelet correlation matrix which is invariant to 
scale shift. This invariance defines a wide-sense self-similarity.
\begin{figure}[h]
\centering
\includegraphics[width=0.7\linewidth]{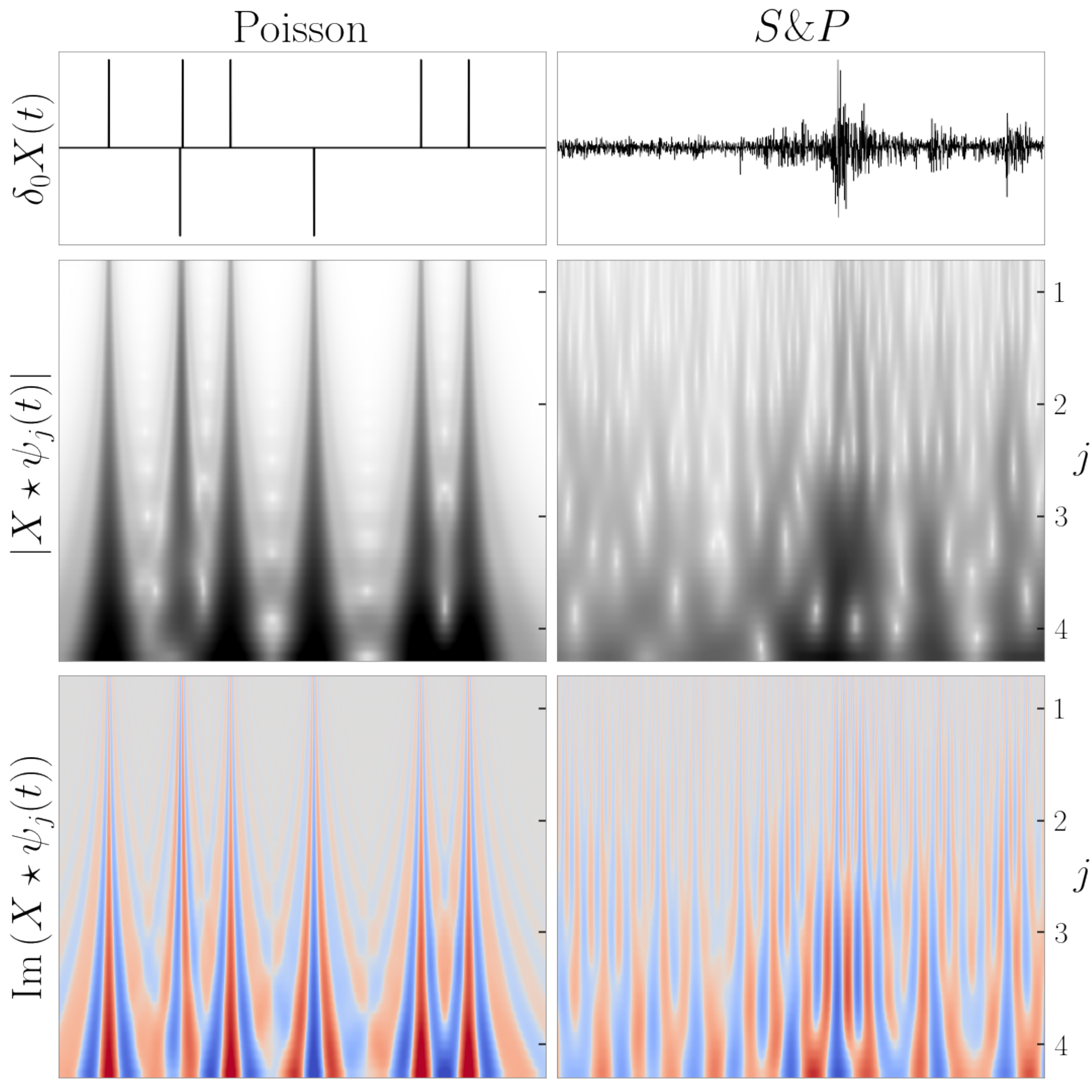}
\caption{Top: increments $\delta_0 X(t)$ of a signed Poisson process and the financial S\&P daily increments from 03/01/2000 to 10/10/2018. Middle: wavelet modulus $|X\star\psi_j(t)|$. The vertical axis corresponds to the log-scale index $j$ which is real. Dark color represents large values.
These modulus have
dependencies across scales produced by Diracs or bursts of activity. Bottom: imaginary part of $X \star \psi_j (t)$. Red and blue colors represent negative and positive values respectively. It shows that 
localized structures
such as Dirac create correlation of phases across octaves, when $j$ increases by $1$ or more.}
\label{FigScalogram}
\end{figure}

\subsection{Scale Dependencies as a Trace of Non-Gaussianity}
\label{SubSecWaveletCovariance}

Section \ref{SubSecWavelets} showed that if $X$ has stationary increments then 
$X \star \psi_j(t)$ and $X \star \psi_{j'}(t')$ are nearly uncorrelated if $|j-j'| > 1.$
If $X$ is Gaussian then these coefficients are jointly Gaussian so it
implies that they are independent. On the contrary, we will now see that 
non-Gaussian time-series
exhibits crucial dependency across
scales. This dependency provides important information on non-Gaussian properties of $X$.

Figure \ref{FigScalogram} shows the wavelet transform of S\&P financial signal, and of a Poisson process whose increments have a random sign. They are calculated with the complex Battle-Lemarié wavelet. Diracs and bursts of activity in the financial signal create high amplitude wavelet coefficients, which propagate across scales. It induces strong dependencies between wavelet modulus across scales. These dependencies also appear in the wavelet transform phase. Diracs produce
high amplitude wavelet coefficients whose phase propagates regularly across octaves, when $j$ increases by $1$ or more. On the contrary, financial bursts of activity have a phase that is randomly modified from one octave to the next. Correlations when $j$ increases by less than $1$ are due to correlations between the wavelets themselves.

To understand this dependency phenomenon, consider a localized
pattern $f(t)$ in the neighborhood of $t=0$, such as a Dirac.
It is randomly translated to define $X(t) = f(t-U)$,
where $U$ is a random variable uniformly distributed in $[0,1]$. Its wavelet coefficients $X \star \psi_j (t) = f \star \psi_j (t-U)$ are centered at $t=U$ at all scales $2^j$, and are thus highly dependent. Their amplitude 
and phase are a signature of the translated pattern $f$. It illustrates the importance
of wavelet coefficient dependencies across scales, for non-Gaussian processes.

\subsection{Joint Phase-Modulus Correlations Across Scales}
\label{SubSecJointWaveletModulus}

This section introduces a joint wavelet phase-modulus correlation matrix which
captures dependencies of wavelet coefficients across scales.
Complex wavelet coefficients have a negligible correlation at different scales because
they are supported in different frequency bands. Their correlation is thus canceled
by phase fluctuations which occur at different rates.
We first realign their frequency support with a modulus, and then compute their correlation.
Correlations of wavelet coefficient modulus were first studied by Portilla and Simoncelli \cite{portilla2000parametric}.
Their properties are analysed in \cite{mallat2020phase,zhang2021maximum}. The joint phase-modulus correlation matrix also preserves phase information, by correlating
wavelet coefficients with and without phases. It partly characterizes
phase alignments across scales.

\subsubsection{Non-zero correlations by removing complex phase}
\begin{figure}
\centering
\includegraphics[width=0.8\linewidth]{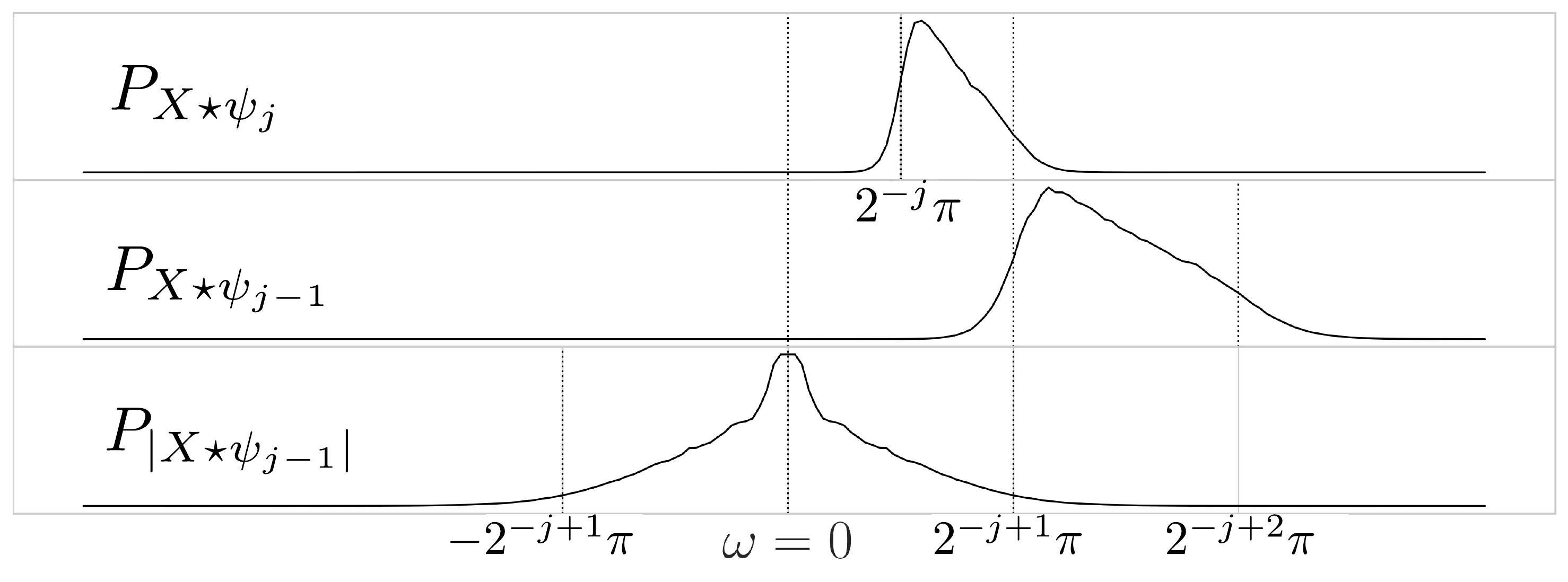}
\caption{Top: power spectrum of $X \star \psi_j$ for the S\&P time-series. It is mostly concentrated in $[2^{-j} \pi, 2^{-j+1} \pi]$. Middle: power spectrum of  $X \star \psi_{j-1}$. Bottom: power spectrum of $|X \star \psi_{j-1}|$ is mostly concentrated in $[-2^{-j+1} \pi, 2^{-j+1} \pi]$ and strongly overlaps with the power spectrum of $X\star\psi_j$.}
\label{FigSupportRealign}
\end{figure}

Eliminating the phase with a modulus can introduce correlations across scales. Indeed,
let us remind that
the cross-spectrum $P_{Y,Z}(\om)$ of two jointly stationary random processes $Y(t)$ and $Z(t)$ is the
Fourier transform of their cross-correlation
\begin{equation}
\nonumber
P_{Y,Z}(\om) = \int \E\{Y(t) Z(t-\tau)^*\} \,e^{-i\tau\om}\, d \tau ,
\end{equation}
and the Cauchy-Schwarz inequality proves that 
\begin{equation}
\label{crossSpectrum}
|P_{Y,Z}(\om)|^2 \leq P_Y (\om)\, P_Z(\om) .
\end{equation}
The cross-correlation of $Y(t)$ and $Z(t-\tau)$ is therefore zero if their power spectra do not overlap.
Applied to $Y = X \star \psi_j$ and $Z = X \star \psi_{j-a}$, it verifies once again that they are essentially uncorrelated if $a \neq 0$. Indeed, their power spectrum do not overlap, as illustrated in Figure \ref{FigSupportRealign}.
However, we now show that the power spectrum of $Y = X \star \psi_j$ and $Z = |X \star \psi_{j-a}|$ or of $Y = |X \star \psi_j|$ and $Z = |X \star \psi_{j-a}|$ can overlap. They can thus have non-zero correlations, after suppressing their mean.

The power spectrum  $P_X(\om) |\widehat{\psi}(2^{j-a} \om)|^2$ of $X \star \psi_{j-a}$
has a support mostly
concentrated in $[2^{-j+a} \pi, 2^{-j+a+1} \pi]$. A modulus eliminates the phase 
of $X \star \psi_{j-a}$ which oscillates at the
center frequency $3\times2^{-j+a-1} \pi$. As a consequence, the power spectrum of
$|X \star \psi_{j-a}|$ is centered at $\om = 0$, and its energy is
mostly concentrated in $[-2^{-j+a} \pi , 2^{-j+a} \pi]$ \cite{mallat2020phase,zhang2021maximum}.
This is shown by Figure \ref{FigSupportRealign}.
It results that the spectrum of $X \star \psi_j$ and
$|X \star \psi_{j-a}|$ do overlap if $a > 0$, and the spectrum of $|X \star \psi_j|$ and $|X \star \psi_{j-a}|$ overlap for any $a$
since they are both centered at $\om = 0$.

\subsubsection{Joint phase-modulus correlation matrix}
Let us write $\rho(z)=(z,|z|)$ for any $z \in \C$. 
We now show how to represent the dependencies of wavelet coefficients
from joint phase-modulus correlation matrix of
\[
\rho WX = (W X , |W X|) = \Big( X \star \psi_j (t) , | X \star \psi_j (t)| \Big)_{t,j} .
\]
If $X$ is sampled at intervals
normalized to $1$ and is of size $N$ then we compute wavelet coefficients
over $\log_2 N$ scales $1 < 2^j \leq N$ corresponding to $1 \leq j \leq \log_2 N$.
If $X$ is stationary then
$\E\{\rho W X\} = (\E\{WX\}\,,\,\E\{|WX|\})$ does not depend upon $t$.
Without loss of generality we suppose
that $\E\{X(t)\} = 0$ so $\E\{W X(t)\} = 0$. 

The coefficients of the joint phase-modulus correlation matrix $\E\{\rho WX \, (\rho WX)^*\}$ are
\[
\Big( \E \{ \rho Wx(t,j) \, \rho Wx(t-2^j \tau,j-a)^* \} \Big)_{t,\tau,j,a} .
\]
They do not depend upon $t$ because $X$ is stationary. 
We estimate them from a single realization of $X$ with a time average
\begin{equation}
\nonumber
\Av{t < N} \big( \rho Wx(t,j) \, \rho Wx(t-2^j \tau,j-a)^* \big).
\end{equation}
The correlation matrix $\E\{\rho WX \, (\rho WX)^T\}$ is
composed of four submatrices 
\begin{equation} \nonumber
    \label{JointPhaseModulus}
	\left(
	\begin{array}{cc}
	\E\{W X \,W X^T\} & \E\{W X \, |W X|^T\}
	\\
	\E\{|W X|\, W X^T\} & \E\{|W X|\, |W X|^T\}
	\end{array}
	\right)
\end{equation}
We show that $E\{W X \,W X^T\}$ and $\E\{W X \, |W X|^T\}$ are sparse matrices. 
On the other hand,
we shall see
that $\E\{|W X|\, |W X|^T\}$ may not be sparse. 
\begin{figure}
\centering
\begin{subfigure}[b]{0.25\linewidth}
    \centering
    \includegraphics[width=\linewidth]{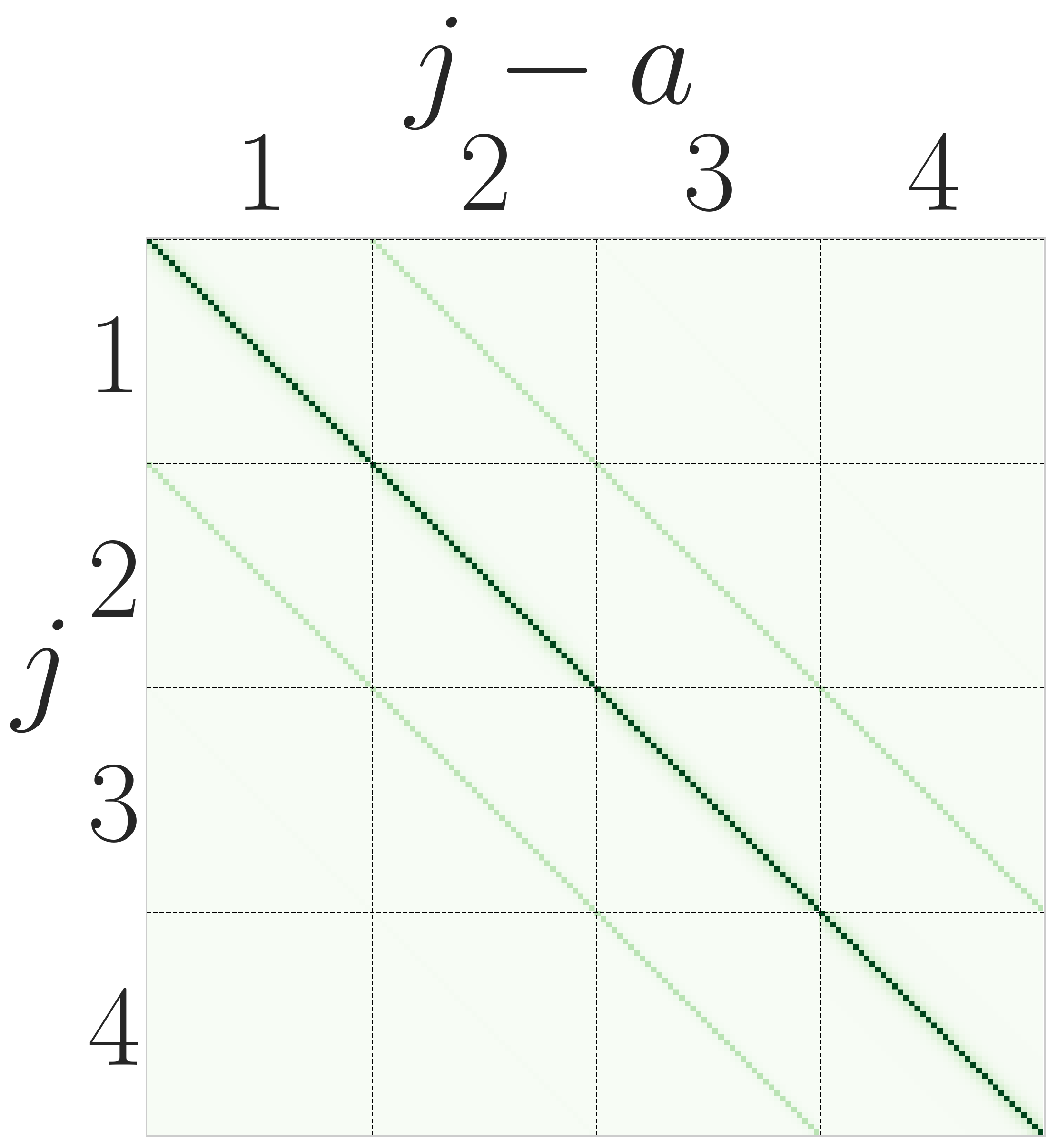}
    \caption{$\E\{W X \,W X^T\}$}
    \label{FigCW}
\end{subfigure}
\begin{subfigure}[b]{0.25\linewidth}
    \centering
    \includegraphics[width=\linewidth]{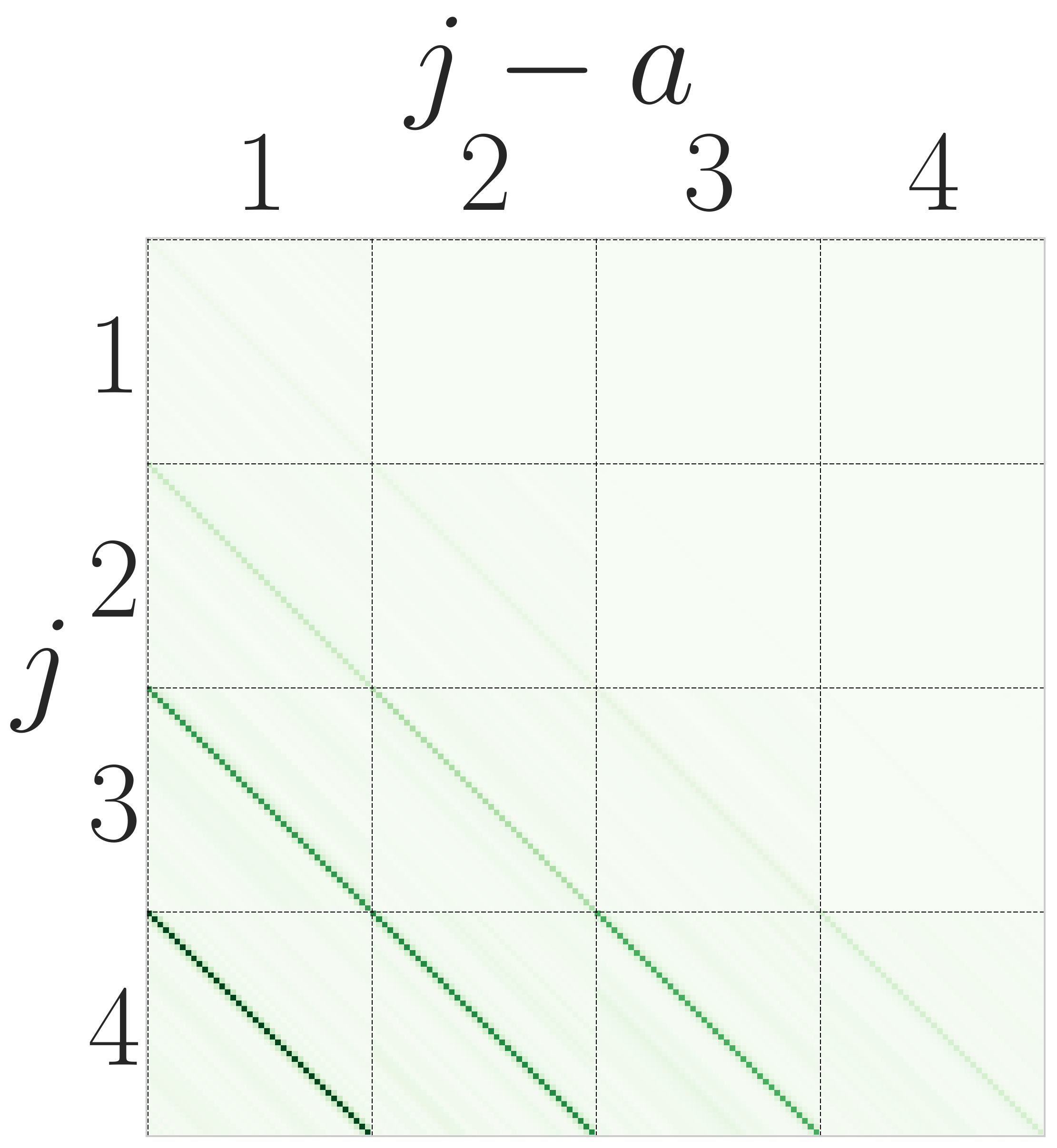}
    \caption{$\E\{W X \, |W X|^T\}$}
    \label{FigCWmW}
\end{subfigure}
\begin{subfigure}[b]{0.25\linewidth}
    \centering
    \includegraphics[width=\linewidth]{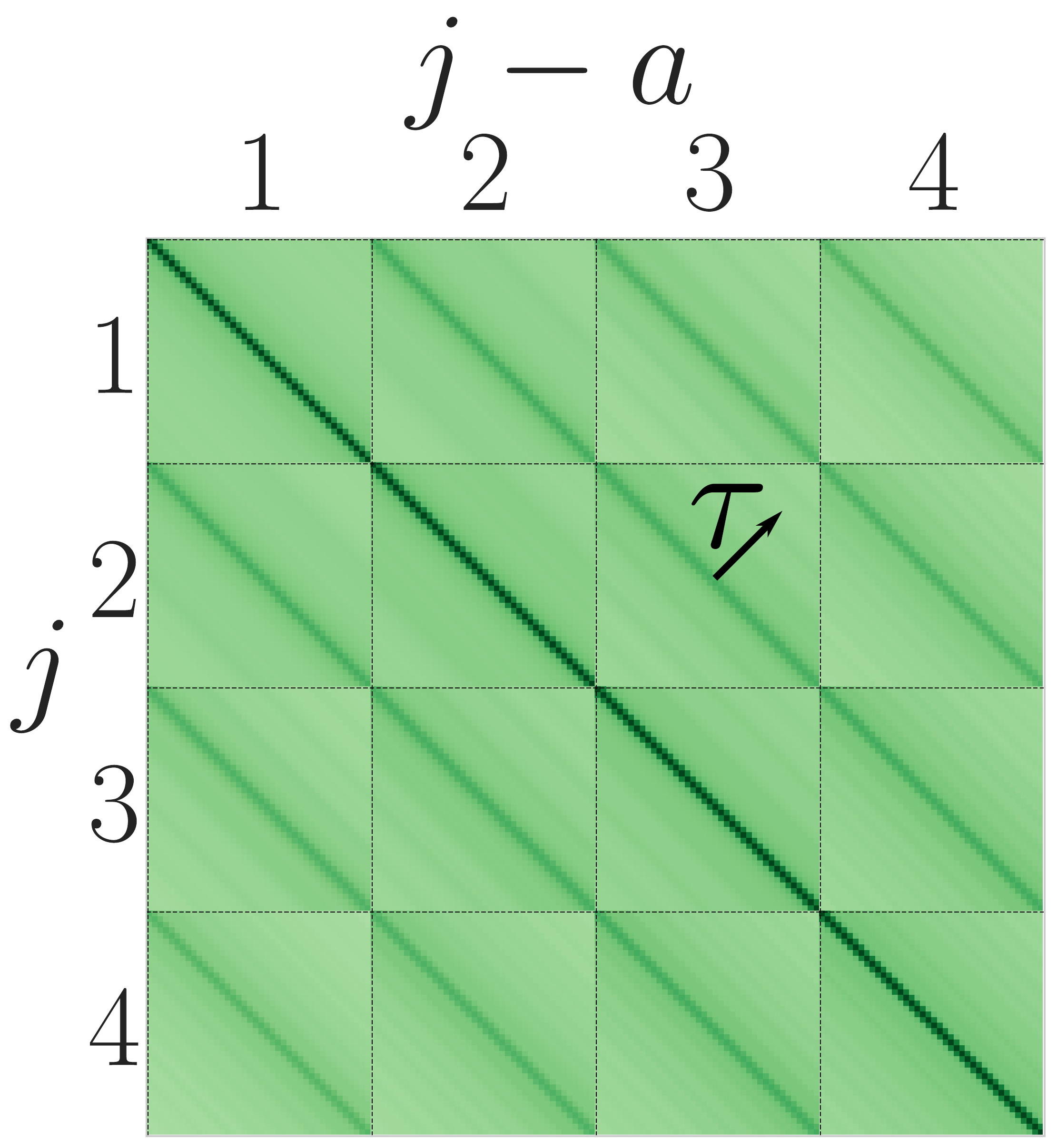}
    \caption{$\E\{|W X| \, |W X|^T\}$}
    \label{FigCmWmW}
\end{subfigure}
\caption{Modulus of the joint phase-modulus correlation
$\E\{\rho WX \, (\rho WX)^*\}$
for S\&P signal. The diagonal of this matrix ($\tau=0,a=0$) is the wavelet spectrum $\sigma^2_W(j)$ which is not constant. To remove such normalization effect, we plot the matrix where each coefficient $(t,\tau,j,a)$ is divided by $\sigma_W(j)\sigma_W(j-a)$.
For the 3 matrices,
each subblock is a Toeplitz correlation matrix along time $(t,t-2^j \tau)$, for scales $(j,j-a)$ fixed, because of time stationarity. All correlation values are constant when $j$ varies, which is a mark of wide-sense self-similarity.}
\label{FigC}
\end{figure}
The coefficients of the wavelet auto-correlation matrix $\E\{W X \,W X^T\}$ are
\begin{equation}
\label{wavelet-autocor}
    \E\{X\star\psi_j(t) \, X\star\psi_{j-a}(t-2^j\tau)^*\}.
\end{equation}
Section \ref{SubSecWavelets} explains that we can approximate it by its diagonal values which define the wavelet spectrum.
For a signal of size $N$, since $1 \leq j \leq \log_2 N$ we only estimate
$\log_2 N$ wavelet spectrum coefficients  $\E\{|X\star\psi_j(t)|^2\}$.

The off-diagonal matrix $\E\{  W X  |W X|^T\}$ is a correlation between complex wavelet coefficients
and their modulus, that we shall call {\it phase-modulus correlation}.
\begin{equation}
\label{wavelet-crosscor}
    \E\{X\star\psi_j(t) \, |X\star\psi_{j-a}(t-2^j\tau)|\}
\end{equation}
Since $\E\{X \star \psi_j\} = 0$ these correlations are also covariance coefficients.
Figure \ref{FigCWmW} shows that wavelet phase-modulus correlations are non-negligible for a scale shift $a \geq 0$ and $\tau = 0$. This is expected because
the power spectrum of $X \star \psi_{j}$ and $|X \star \psi_{j-a}|$ have an overlapping support. When $\tau \neq 0$ they become negligible because of random phase fluctuations.
Since $1 \leq j-a < j \leq \log_2 N$, we only estimate
$2^{-1} \log_2 ^2 N$ wavelet phase-modulus cross-spectrum coefficients
$\E\{X\star\psi_j(t) \, |X\star\psi_{j-a}(t)|\}$.

The coefficients of the wavelet {\it modulus auto-correlation} $\E\{|W X|\,|W X|^T\}$ are
\begin{equation}
\label{waveletmod-autocor}
    \E\{|X\star\psi_j(t)| \, |X\star\psi_{j-a}(t-2^j\tau)|\}.
\end{equation}
Figure \ref{FigCmWmW} shows that the wavelet modulus correlation is nearly a full
matrix. Their covariance is obtained by subtracting the modulus means. 
These covariances 
are also a priori non-zero for all scale shift $a$, because the power spectra
of $|X \star \psi_j|$ and $|X \star \psi_{j-a}|$ have overlapping supports. They can
be non-negligible for large time shift $\tau$, 
because all phases have been eliminated. The number of time shifts is nearly equal to the
signal size $N$. For $1 \leq j-a \leq a \leq \log_2 N$, there are 
about $2^{-1}\, N\, \log^2_2 N $ potentially non-negligible wavelet modulus correlations,
which is too large to estimate them directly from a single realization of $X$.
Section \ref{SubSecScatteringCovariance} shows that one can reduce the number of correlation coefficients 
to $\log_2^3 N$, by applying a second wavelet transform on $|X \star \psi_j (t)|$ before
calculating its auto-correlation.

\subsubsection{Non-Gaussian properties}
Phase and modulus wavelet correlations can be analyzed as particular cases
of phase harmonic correlations introduced in \cite{mallat2020phase}.
It captures non-Gaussian properties proved in \cite{zhang2021maximum}.
The following proposition transpose these results in our context. 
It proves that the existence of non-negligible 
wavelet modulus covariances across scales is a mark of non-Gaussianity.
We write $\Cov\{A , B\} = \E\{A B^*\} - \E\{A\}\,\E\{B\}^*.$
\begin{proposition}
\label{prop:non-gaussian-cov}
If $X$ is Gaussian and $\widehat{\psi}_j\, \widehat{\psi}_{j-a} = 0$ then for all $\tau$
\[
\Cov \{ \rho Wx(t,j) \,,\, \rho Wx(t-2^j \tau,j-a) \} =  0.
\]
\end{proposition}
We indeed saw that if  $\widehat{\psi}_j\widehat{\psi}_{j-a}=0$ then
$X \star \psi_j (t)$ and $X \star \psi_{j-a}(t-2^j\tau)$
are uncorrelated (\ref{eq:wx-correlation}). If $X$ is Gaussian then they are also Gaussian and
hence independent. Applying a modulus preserves this independence and thus produces covariance coefficients which remain zero, proving the second equality. 
The condition $\widehat{\psi}_j\, \widehat{\psi}_{j-a} = 0$
is verified up to a very small error for $|a| > 1$. For $|a| = 1$,
the supports of $\widehat{\psi}_j$ and $\widehat{\psi}_{j-a}$ have a small overlap so the product is small but not zero.
It follows from the proposition that non-zero covariance coefficients across distant scales evidence that 
$X$ is not Gaussian.
 
Time-asymmetry
is another form of non-Gaussianity, often produced by
causality phenomena.
Let $R$ be the time reversal operator $Rx(t)=x(-t)$. A process $X$ is
said to be time-reversible if the probability distributions of $R X$ and $X$ are equal. Gaussian stationary processes are time-reversible.
The following proposition shows that
time-reversibility can be detected from phase correlation coefficients.
\begin{proposition}
\label{PropTimeReversal1}
If $X$ is time-reversible then the joint wavelet modulus correlation has a Hermitian symmetry along $\tau$:
\begin{equation}
\label{propeasdf}
\E \{ \rho WX(t,j) \, \rho WX(t-2^j \tau,j-a)^* \} = 
\E \{ \rho WX(t,j)^* \, \rho WX(t + 2^j \tau,j-a) \} .
\end{equation}
\end{proposition}
Time-reversibility means that $X$ and $RX$ have the same distribution. Since
$(RX)\star\psi_j(t)=X\star\psi_j(-t)^*$ it implies the equality (\ref{propeasdf}).
This Hermitian symmetry is always satisfied by the wavelet auto-correlation coefficients 
(\ref{wavelet-autocor})
even if $X$ is not time-reversible, but not necessarily
by phase-modulus correlations and modulus auto-correlation coefficients (\ref{wavelet-crosscor},\ref{waveletmod-autocor}).
If they do not have the Hermitian symmetry (\ref{propeasdf})
then $X$ is not time-reversible, and hence non-Gaussian.

\subsection{Wide-Sense Self-Similarity}
\label{SubSecWideSenseSelfSimilarity}

Self-similarity is defined in (\ref{strongSelfSimilaritydX})
and (\ref{strongSelfSimilarityWX}) 
on the process distributions, which are high-dimensional objects,
on increments or wavelet coefficients. Such properties cannot be 
tested statistically on a single realization.
Section \ref{SecMarginalMoments} gives necessary conditions (\ref{marginalSelfSimilaritydX},\ref{marginalSelfSimilarityWX}) over the marginals of
increments and wavelet coefficients, which are simple to verify but 
provides relatively weak characterization of self-similarity.
The same difficulty appears to test that a process has a stationary distribution. It cannot be tested statistically 
on a single realization.
Conditions on marginals
impose that the probability distributions of $X(t)$ for each $t$ does not
depend upon $t$. It can be tested numerically but it is a weak condition.
More powerful characterizations of stationarity impose that $\E\{X(t)\}$ and
the auto-correlation of $X$ are invariant to time shift. The process $X$ is then said to be wide-sense stationary. We follow the same approach for self-similarity by imposing
a scale shift invariance on a normalized joint phase-modulus correlation matrix. 

Self-similarity of increments distributions (\ref{strongSelfSimilaritydX}) or
wavelet coefficients (\ref{strongSelfSimilarityWX}) are defined up to random multiplicative
factors $A_\ell$. We eliminate these multiplicative factors 
with a 
normalized
phase-modulus correlation matrix where each correlation coefficient of
$\E\{\rho WX \, (\rho WX)^*\}$ is normalized by a product of standard deviations
given by $\sigma_W^2 = \E \{|X \star \psi_j (t)|^2\}$:
\begin{equation} \nonumber
    C_{\rho W}(\tau;j,a) = \frac{    \E \{ \rho WX(t,j) \, \rho WX(t-2^j \tau,j-a)^* \}}
    {\sigma_W(j)\, \sigma_W(j-a)} .
\end{equation}

A wavelet transform introduces explicitly
a scaling parameter $2^j$. However, 
the correlations of wavelet coefficients $WX$
vanish across scales, and thus can not
be used directly to identify self-similarity across scales.
We define a notion of wide-sense self-similarity as a
translation and scale invariance of
the mean and correlation matrix of $(W X , |W X|)$.
\begin{theorem}[Wide-sense self-similarity]
\label{main-theo}
If $X$ is self-similar up to the scale $2^J$ in the sense of (\ref{strongSelfSimilarityWX}) 
then there exist $c_1$, $c_2$, $\zeta_1$, $\zeta_2$ such that for all $j \leq J$
\begin{equation}
\label{DefOrder1}
\E\{|X \star \psi_j (t)|\}  =  c_1\, 2^{j \zeta_1} ,
\end{equation}
\begin{equation}
\label{DefOrder2}
\E\{|X \star \psi_j (t)|^2\}  =  c_2\, 2^{j \zeta_2} .
\end{equation}	
and for all $\tau$, $j \leq J$, $a$,
\begin{eqnarray}
    \label{DefCross}
    C_{\rho W}(\tau; j, a) & = & C_{\rho W}(\tau; 0, a).
\end{eqnarray}
\end{theorem}
The theorem is proved in \ref{AppendixProofTheorem}.
A process $X$ which satisfies the properties (\ref{DefOrder1},\ref{DefOrder2},\ref{DefCross}) is said to be
wide-sense self-similar. The appendix proves that self-similarity implies wide-sense 
similarity. Similarly to moment self-similarity 
(\ref{marginalSelfSimilarityWX}), 
wide-sense self-similarity imposes the existence of scaling exponents $\zeta_q$, but only for $q = 1$ and $q = 2$.
The scaling exponent $\zeta_2$ specifies the decay of the wavelet spectrum
$\sigma_W (j) =   \E\{|X \star \psi_j (t)|^2\}$.
The ratio between first and second order moments 
is a sparsity measure on wavelet coefficients
\begin{equation}
\label{sparsityCoeff}
s_W(j) =  \frac {\E\{|X\star\psi_j(t)|\}} {\sigma_W (j) }.
\end{equation}
If $X$ is wide-sense self-similar then $s_W^2(j)=c_s\, 2^{j \zeta_s} \leq 1,$
where $c_s = c_1^2\,c_2^{-1}$ and $\zeta_s = 2 \zeta_1 - \zeta_2\geq0$. The lower $s_W^2(j)$ the sparser $X\star\psi_j(t)$.
The exponent $\zeta_s$ is a sparsity rate
which governs the increase of sparsity when the scale decreases.
If $X$ is Gaussian then $ \zeta_s = 0$.
If $\zeta_s > 0$ then the sparsity
of $X \star \psi_j$ increases as $j$ decreases.
The constant $c_s $ is a sparsity multiplicative factor.
If $X$ is Gaussian then $X \star \psi_j$ is also Gaussian and one can verify that 
$c_s = \pi / 4 $, which is the ratio between first and second order moments of complex Gaussian random variables.

Wide-sense self-similarity also imposes
that the normalized phase-modulus correlation matrix $C_{\rho W}$ depends only on time shift $\tau$ and scale shift $a$. This is a powerful second order condition, which is sufficient to reveal existence of 
important
self-similar properties in time-series.
It applies to the non-negligible coefficients of each of the three
submatrices of $C_{\rho W}$. 
Over diagonal wavelet auto-correlation coefficients, it is already
specified by (\ref{DefOrder2}). Over non-negligible
phase-modulus cross-spectrum coefficients, it imposes that
\begin{equation} \nonumber
    \label{phase-mod-cross-spe}
    C_{W|W|}(0;j,a) = \frac{\E\{ WX(t,j) \, |WX(t,j-a)| \}}
    {\sigma_W(j)\, \sigma_W(j-a)}
\end{equation}
does not depend upon $j$. These coefficients are estimated by replacing each expected value
by an average on $t$. For self-similar processes, since these moment do not depend on $j$, we can further improve this estimation by averaging them over scales. It defines
an {\it scale invariant phase-modulus cross spectrum}
\begin{equation}
\label{phase-mod-cross-spe-av}
\overline C_{W|W|}(a) = \Av{j} C_{W|W|}(0;j,a) .
\end{equation}
For wavelet modulus auto-correlations (\ref{waveletmod-autocor}), these conditions are translated into conditions
over a scattering cross-spectrum which is introduced in the next section.

Processes such as fractional Brownian motion \cite{mandelbrot1968fractional} or multifractal random walk \cite{MRW01} are self-similar and therefore wide-sense self-similar. Self-similarity cannot be
tested statistically,
whereas wide-sense self-similarity is a correlation property which can be estimated. 
Figure \ref{FigC} shows that the S\&P financial signal is wide-sense self-similar. Indeed
$\log \E\{|X \star \psi_j|\}$ and $\log \E\{|X \star \psi_j|^2\}$ have a linear decay
along $j$, and the normalized correlation $C_{\rho W}$ is constant along its diagonals when $j$ varies.

\section{Scattering Cross-Spectrum}
\label{SubSecScatteringCovariance}

Section \ref{SubSecJointWaveletModulus} explains that there are too many wavelet modulus
auto-correlation coefficients to estimate them from a single realization of $X$.
We introduce a low-dimensional approximation of this auto-correlation matrix
by applying a second wavelet transform, which defines a scattering transform \cite{mallat2012group}. The resulting scattering covariance
is nearly diagonal and its spectra can be estimated from a single realization.

\subsection{Diagonal Scattering Cross-Correlation}
\label{SubSecScatteringCrossSpectrum}

Section \ref{SubSecWavelets} explains
that if $X$ has self-similarity properties then its auto-correlation matrix is well approximated by a diagonal matrix after applying a wavelet transform. 
Similarly, instead of computing directly the auto-correlation of
$|WX| =\big(|X \star \psi_j(t)|\big)_{j,t}$ we will compute the auto-correlation
of its wavelet transform.

Applying a second wavelet transform $W$ on $|W X|$ defines a scattering transform
$S X = W \, |W X|$, with
\begin{equation}
\nonumber
S X (t;j,k) := |X\star\psi_{j}|\star\psi_{k}(t) .
\end{equation}
It is non-negligible only if $k > j$. Indeed, the Fourier transform of $|X \star \psi_{j}|$ is mostly concentrated in $[-2^{-j} \pi, 2^{-j} \pi]$. If $k \leq j$ then it does not intersect the frequency interval $[2^{-k} \pi, 2^{-k+1} \pi]$ where the energy of
$\widehat{\psi}_k$ is mostly concentrated, in which case
$S X(t;j,k) \approx 0$. 

Since $S X = W \, |W X|$, its auto-correlation is
\begin{equation}
    \label{scatProjection}
    \E\{S X\,S X^T\} = W\,\E\{|W X|\,|W X|^T \}\, W^{T} ,
\end{equation}
which specifies $\E\{|W X|\,|W X| \}$ because $W$ is invertible. 
The coefficients of $\E\{ S X\, S X^T\}$ are
\[
\E\{|X\star\psi_{j}|\star\psi_{k}(t) \, |X\star\psi_{j-a}|\star\psi_{k'}(t-\tau)^*\}
\] 
for all time $t$, time shift $\tau$, first wavelet scale $j$ and scale shift $a$, and second wavelet scales $k,k'$. We impose that $k > j$ and $k' > j-a$ otherwise
the scattering coefficients are negligible.
Applying (\ref{crossSpectrum}) to $Y(t) = |X\star\psi_{j}|\star\psi_k(t)$ and
$Z = |X\star\psi_{j-a}|\star\psi_{k'}(t-\tau)$ shows that this
correlation is negligible if $k \neq k'$ because the spectra of $Y$ and $Z$ barely overlap. Indeed
$\widehat{\psi}_k$ and $\widehat{\psi}_{k'}$ are concentrated over non-overlapping frequency intervals. 
Correlations for $k=k'$ have a fast polynomial decay away from $\tau=0$ \cite{wornell1993} when the cross-spectrum of the modulus is regular,
and we thus shall only 
consider
these scattering
correlations for $\tau = 0$. Scattering correlations are thus calculated only
for $k=k'=j-b$ with $b < 0$ and $\tau = 0$.

We incorporate the normalization of the wavelet modulus auto-correlation by
the wavelet spectrum
$\E\{|X \star \psi_j|^2\}$ 
to the scattering correlations.
The normalized diagonal scattering coefficients for $k = k' = j-b$ define 
a {\it scattering cross-spectrum} whose coefficients are:
\begin{equation} \nonumber
\label{scat-cross-pect}
    C_S (j,a,b) = \frac{\E\{|X\star\psi_{j}|\star\psi_{j-b}(t) \, |X\star\psi_{j-a}|\star\psi_{j-b}(t)^*\}} 
     {\sigma_W (j)\, \sigma_W(j-a)} .
\end{equation}
Since $\psi_{j-b}$ has a Fourier transform mostly supported in 
$[2^{-j+b} \pi, 2^{-j+b+1} \pi]$, $C_S(j,a,b)$ can be interpreted as
the cross-spectrum of $|X \star \psi_j|$ and $|X \star \psi_{j-a}|$ integrated over this frequency interval.
These cross-spectra specify intermittency phenomena which appear when the wavelet modulus correlations (\ref{waveletmod-autocor}) remain large on a long-range of time. 
If the modulus are uncorrelated in time across scales then
$C_S (j,a,b)$ is nearly constant along $b$ for $j,a$ fixed.
On the contrary, if $C_S (j,a,b)$ has a fast decay in $b$ then it implies that the modulus have long range correlations in time.

If $X$ is of size $N$ then there are at most $\log_2 N$ scales indices $j$, $a$ and $b$. 
it shows that the scattering transform can provide an approximation of
$N \, \log_2^2 N$ wavelet modulus auto-correlation coefficients 
with $\log_2^3 N$ scattering cross-spectrum coefficients.

\subsection{Properties}

The following proposition derives from
Proposition \ref{PropTimeReversal1} that the imaginary part of $C_S$ captures 
time-asymmetry properties of $X$. The proof is in \ref{AppendixProofOfPropositionTimeReversalAndTheorem}.
\begin{proposition}
\label{PropTimeReversal2}
If $X$ is Gaussian and $\widehat{\psi}_j\, \widehat{\psi}_{j-a} = 0$ then
\begin{equation} 
    \nonumber
    C_S (j,a,b) = 0 .
\end{equation}
If $X$ is time-reversible then for all $j$, $a$ and $b$ the imaginary part satisfies
\begin{equation} 
    \nonumber
    {\rm Im} \, C_S(j, a, b) = 0.
\end{equation}
\end{proposition}
The following theorem proves that the self-similarity condition (\ref{DefCross}) on 
$C_{\rho W}$ can be evaluated on non-zero scattering coefficients. 
\begin{theorem}
\label{sndTheorem}
The scale invariance (\ref{DefCross}) of $C_{\rho W}$ implies that
\begin{equation}
    \label{selfSimilarityS}
    \forall j\leq J ~~ , ~~ C_S (j,a,b) =   C_S (0,a,b).
\end{equation}
\end{theorem}
The theorem is proved in \ref{AppendixProofOfPropositionTimeReversalAndTheorem}. This condition on scattering cross-spectrum coefficients is necessary and almost sufficient to guarantee the scale invariance of the normalized wavelet modulus auto-correlation
$\E\{|WX|\,|WX|^T\}$. To do so we would also need to impose an invariance condition on the
off-diagonal coefficients of $C_S$ but these coefficients have mostly a negligible amplitude.
In the following, we shall thus systematically replace $\E\{|WX|\,|WX|^T\}$ by the scattering cross-spectrum correlation $C_S$ and assess the scale invariance from (\ref{selfSimilarityS}).
They are estimated by replacing expected values by a time averaging.
For self-similar processes, $C_s$ does not depend on $j$. These invariant
coefficients are thus estimated by averaging them over scales. It defines
a {\it scale invariant scattering cross-spectrum}
\begin{equation}
\label{scat-cross-spe-av}
\overline C_{S}(a,b) = \Av{j} C_{S}(j,a,b) .
\end{equation}
For a signal of size $N$, there are
at most $\log_2 ^2 N$ such coefficients.

\section{Numerical Dashboard for multi-scale Processes}
\label{SecNumericalDashboard}

We show that the scattering spectra,
defined as the multi-scale moments (\ref{wavepsecnfsd},\ref{sparsityCoeff},\ref{phase-mod-cross-spe-av},\ref{scat-cross-spe-av}), 
specify
intermittency, time-asymmetries and self-similar properties
of multi-scale processes.
Next section studies standard mathematical models of multi-scale processes, and
the following section considers numerical financial and turbulence time-series.

\subsection{Models of self-similar processes}
\label{SubSecModelsOfSelfSimilarProcesses}

\begin{figure}
\centering
\begin{subfigure}[b]{0.33\linewidth}
    \centering
    \includegraphics[width=\linewidth]{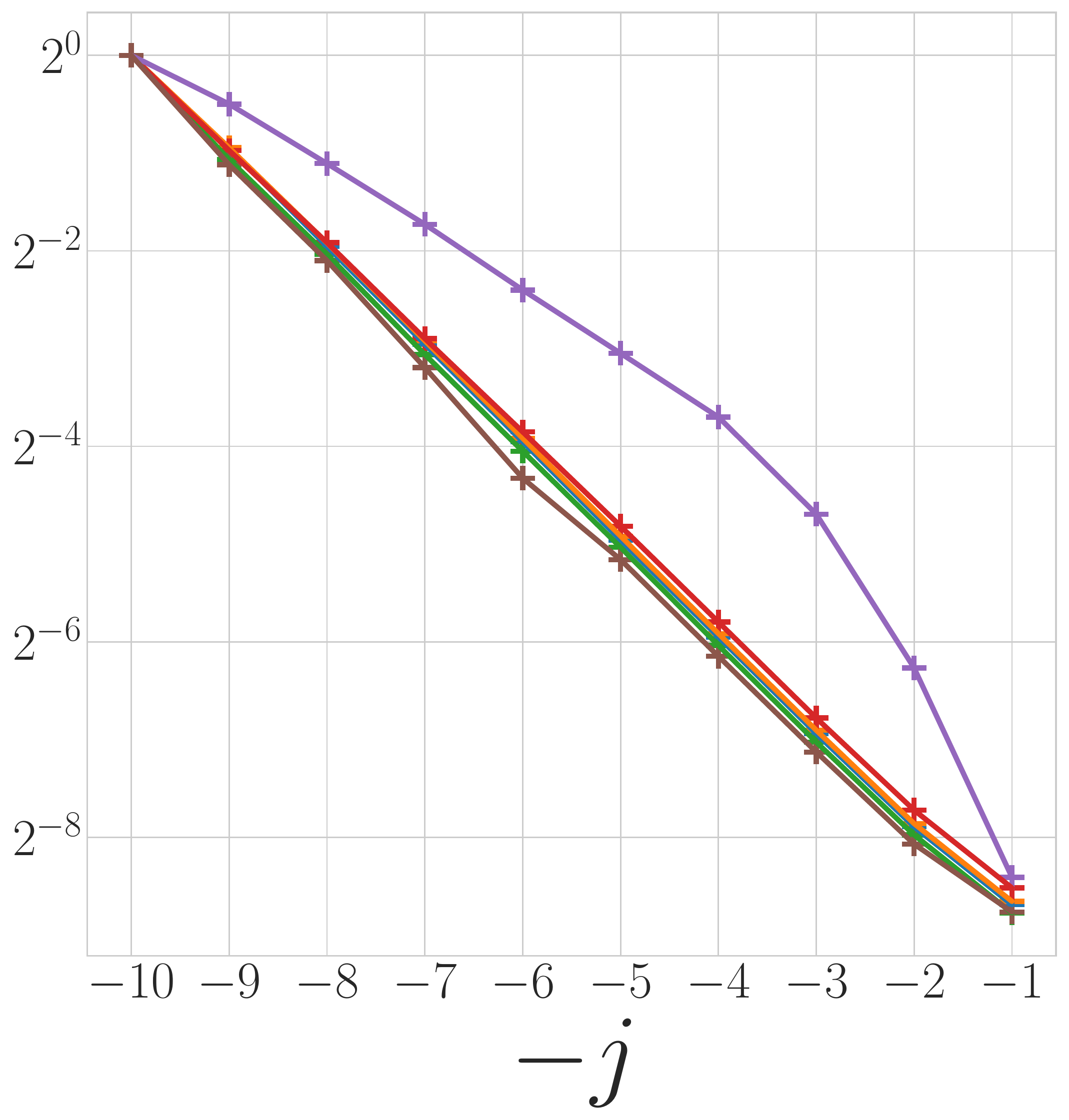}
    \caption{Wavelet power spectrum $\sigma_W^2$}
    \label{FigWaveletSpectrum}
\end{subfigure}
\begin{subfigure}[b]{0.33\linewidth}
    \centering
    \includegraphics[width=\linewidth]{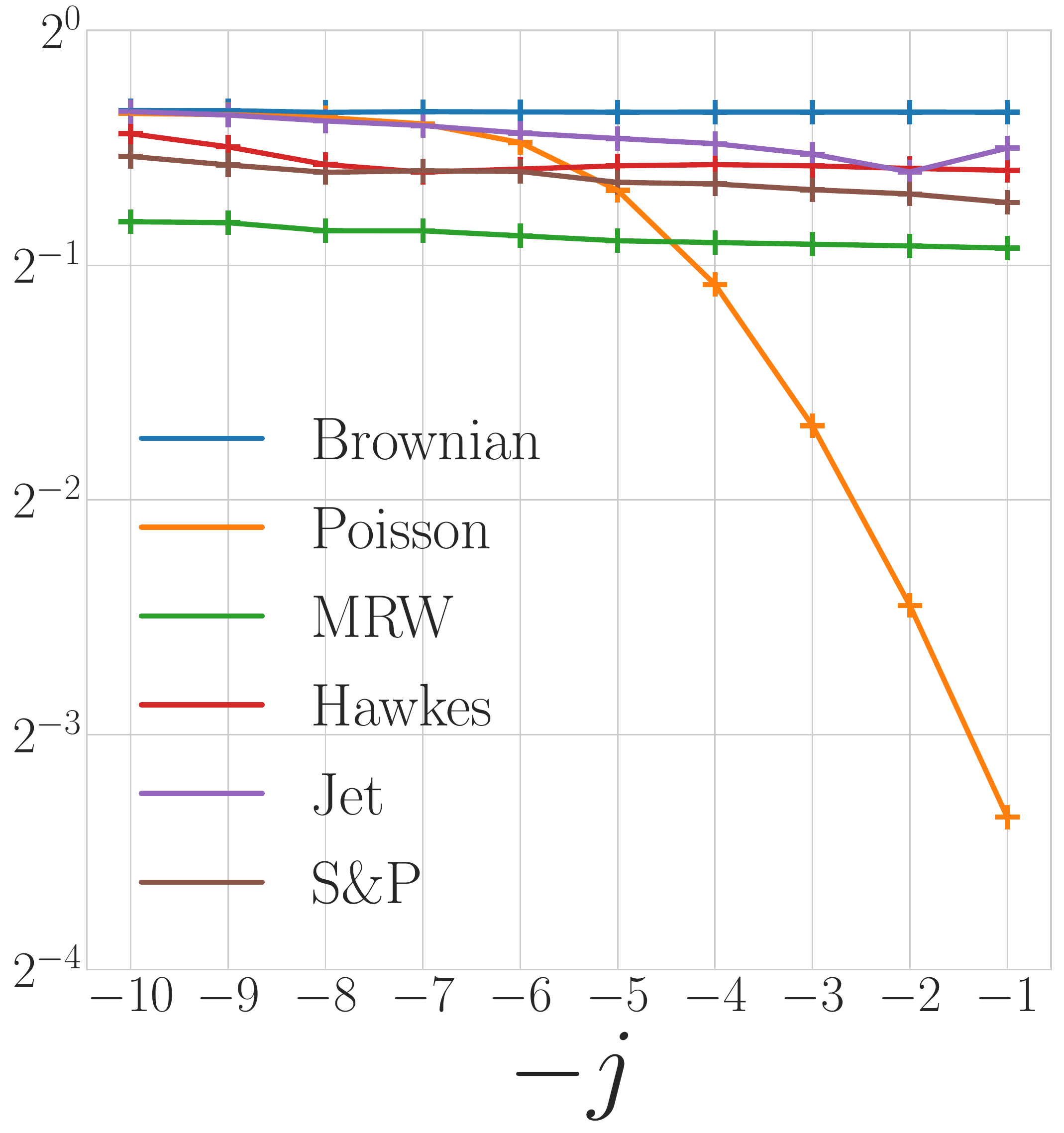}
    \caption{Sparsity factor $s_W^2$}
    \label{FigSparsity}
\end{subfigure}
\caption{(a) Power spectrum $\sigma_W^2(j)$ in (\ref{wavepsecnfsd}) as a function of $-j$, which is a log frequency index. (b) Sparsity factor $s_W^2(j)$ in (\ref{sparsityCoeff}) as a function of $-j$. Each process is shown with a different color. Non-linear curves reveal absence of self-similarity for Poisson process and the turbulent jet at fine scales.}
\label{FigMarginalMoments}
\end{figure}

\begin{figure}[t]
\centering
\includegraphics[width=\linewidth]{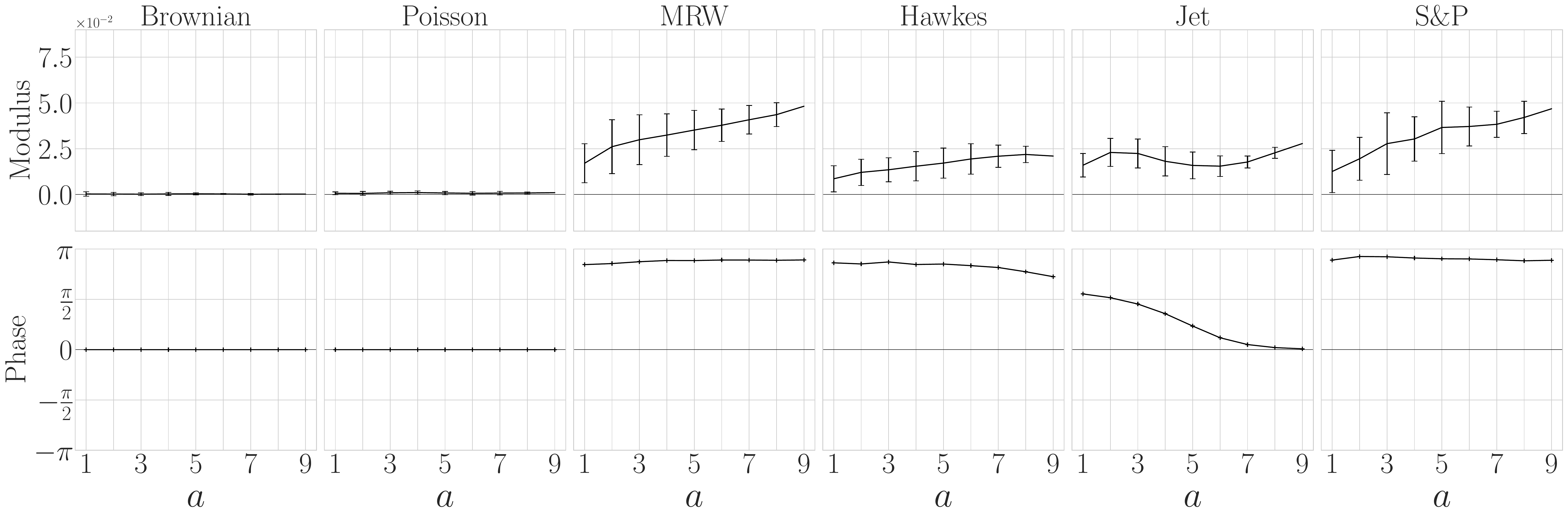}
\caption{Modulus and phase of the scale invariant 
phase-modulus cross-spectrum $\overline C_{W|W|}(a)$ (\ref{phase-mod-cross-spe-av}). 
A skewed MRW and a quadratic Hawkes are shown. 
Error bars represent the 
mean-square variations of $C_{W|W|}(0;j,a)$ around $\overline C_{W|W|}(a)$. Non-zero phase coefficients reveal time-asymmetry.}
\label{FigPhasedEnvelopeCorrelations}
\end{figure}

\begin{figure}[t]
\centering
\includegraphics[width=\linewidth]{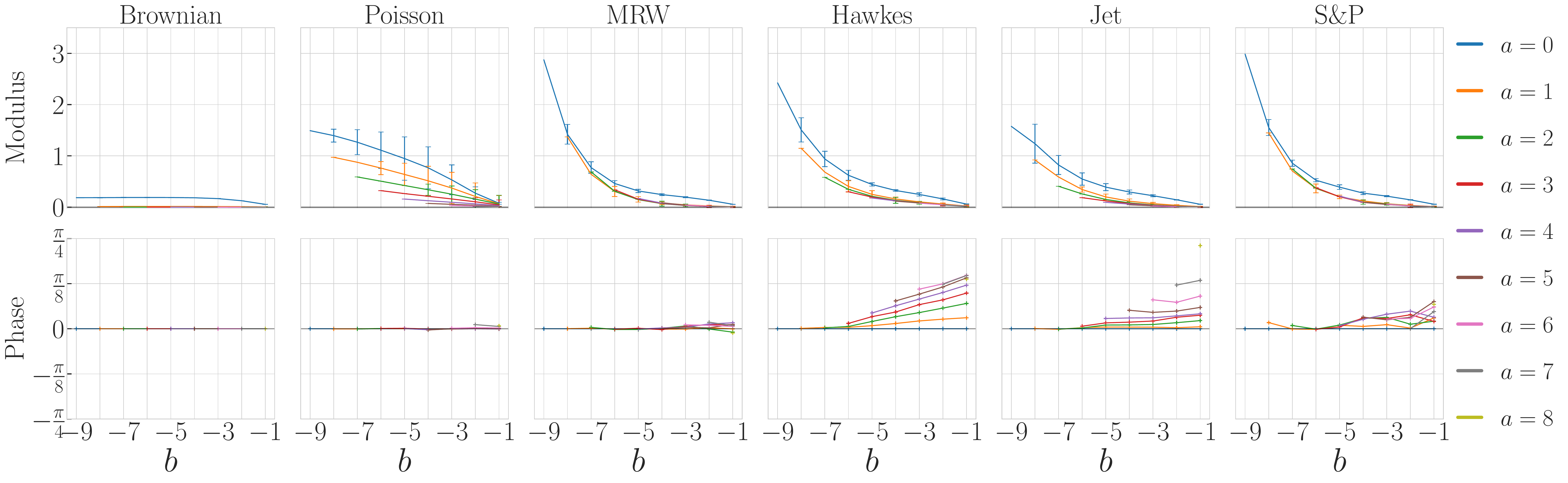}
\caption{Modulus and phase of the scale invariant scattering cross-spectrum $\overline C_S(a, b)$ (\ref{scat-cross-spe-av}) as a function of $b$. 
A skewed MRW and a quadratic Hawkes are shown. 
The parameter $b$ is a log-frequency. Each color curve corresponds to a different scale shift $a$. 
Error bars represent the 
mean-square variations of $C_S(j,a,b)$ around $\overline C_S(a,b)$.
Non-zero phases reveal time-asymmetry of wavelet modulus. }
\label{FigEnvelopeCorrelations}
\end{figure}

We consider Brownian motions, Poisson processes,  multifractal random walks and Hawkes processes.
To analyze their self-similarity properties we
display and analyze their scattering spectra, composed of
\begin{itemize}
\item $\sigma_W^2(j)$: wavelet spectrum (\ref{wavepsecnfsd}) shown in Figure \ref{FigWaveletSpectrum},
\item $s_W^2(j)$: wavelet sparsity factor (\ref{sparsityCoeff}) in Figure \ref{FigSparsity},
\item $\overline C_{W|W|}(a)$: scale invariant phase-modulus cross-spectrum (\ref{phase-mod-cross-spe-av}) in Figure \ref{FigPhasedEnvelopeCorrelations}.
\item $\overline C_S(a,b)$: scale invariant scattering cross-spectrum (\ref{scat-cross-spe-av}) in Figure \ref{FigEnvelopeCorrelations}.
\end{itemize}
Table \ref{TableMarginal12Moments} gives the power-law decay parameters of $\sigma_W^2$ and $s_W^2$. 

Expected values are estimated with empirical averages over $N$ samples in time, as in (\ref{waveletspectestima}). If $X$ has a finite integrable scale $s$, which means that $X(t)$ and $X(t')$ are independent if $|t-t'| > s$ (see section \ref{subsec:power-spectrum}), then we set the maximum wavelet scale $2^J$ to be equal to $s$. When the signal size $N$ goes to $\infty$, time average estimators are  consistent estimators of expected values. Indeed, if the wavelet has a support of size $\alpha$ then all scattering spectra coefficients are
expected values of operators whose support sizes are at most $2 \alpha 2^J$. 
One can thus verify that each empirical estimator averages at least 
$N (2 \alpha s)^{-1}$
blocks of independent coefficients, which have the same mean because $X$ is stationary. 
These empirical estimators thus converge to the expected value when $N$ increases, with a variance which decays at least like $2 \alpha s N^{-1}$.

In this section, we consider multi-scale processes whose integrable scale is not necessarily finite. 
The maximum wavelet scale $2^J$ is chosen to be smaller than the signal size $N$ in order for large scale coefficients to be well estimated.
Time average estimators of scattering spectra coefficients can then provide consistent estimators at all the scales smaller than $2^J$. The variance decay of our estimators is not guaranteed mathematically, but it is verified numerically.
\begin{table}[!h]
\centering
\scalebox{0.8}{
\begin{tabular}{|c||c|c|c|c|c|c|c|}
\hline
& Brown. & Poisson & MRW & Hawkes & Jet & S\&P
\\
\hline
$\zeta_2$ & 1.0 & 1.0 & 1.0 & 1.0 & 0.66 & 1.0
\\
\hline
$c_s$ & 0.79 & $\times$ & 0.66 & 0.65 & 0.67 & 0.61
\\
\hline
$\zeta_s$ & 0.00 & $\times$ & 0.016 & 0.013 & 0.025 & 0.016
\\
\hline
\end{tabular}
}
\caption{Self-similarity parameters $\zeta_2, c_{s},\zeta_s$, 
for different multi-scale processes. For the jet, $\zeta_2, c_s, \zeta_s$ are given on the self-similar range.}
\label{TableMarginal12Moments}
\end{table}

\paragraph{Fractional Brownian motion} It is a Gaussian self-similar  process,
studied by Mandelbrot \cite{mandelbrot1968fractional}. It has stationary increments and a generalized power spectrum
$P_X (\om) = c_2\,|\om|^{-\zeta_2-1}$. Computations are performed with $\zeta_2 = 1$, 
which corresponds to a standard Brownian motion.
Wavelet sparsity coefficients have a power law decay
with $c_s = \pi / 4$ and $\zeta_s = 0$ in Table \ref{TableMarginal12Moments}, because it is a Gaussian process.
Propositions \ref{prop:non-gaussian-cov} and  \ref{PropTimeReversal2} prove that
$\overline C_{W|W|} (a) \approx 0$ and $\overline C_S(a,b) \approx 0$ for $a > 0$, which is verified in Figure \ref{FigPhasedEnvelopeCorrelations} and
Figure \ref{FigEnvelopeCorrelations}. For $a=0$, 
we also observe in Figure \ref{FigEnvelopeCorrelations} that
$\overline C_S(0,b)$ is constant, which shows that  $|X\star\psi_j(t)|$ are uncorrelated at
time increments which are sufficiently large relatively to the scale $2^j$.
As proved by propositions \ref{PropTimeReversal1} and \ref{PropTimeReversal2},
the phases of $C_{W|W|}$ and $C_S$ are zero in Figures \ref{FigPhasedEnvelopeCorrelations} and \ref{FigEnvelopeCorrelations},
because a Gaussian process is time-reversible.
Since Brownian motions are self-similar, the phase-modulus cross-spectrum $C_{W|W|}$ and the scattering cross-spectrum $C_S$ remain constant across scales $2^j$.

\paragraph{Poisson process} It is a jump process having independent stationary increments. The number of jumps in an interval is proportional to the intensity $\lambda$. We further suppose that each jump is positive or negative
with a probability $1/2$. Its power spectrum has a power law decay with $\zeta_2 = 1$, but it is non-Gaussian and not self similar, which clearly appears in its scattering spectra.
Figure \ref{FigSparsity} shows that $\log s_W^2(j)$ in (\ref{sparsityFactor}) has a slope which varies 
as a function of $-j$.
Indeed, if $2^{j} \ll \lambda^{-1}$ then  \cite{bruna2015intermittent} proves that
$s_W^2(j) \sim 2^j$ because
\[
\E\{|X\star\psi_j(t)|\}^2 \sim \lambda^2\,2^{2j}~~\mbox{and}~~ \E\{|X\star\psi_j(t)|^2\} \sim \lambda^2\, 2^j,
\]
whereas if $2^j \gg \lambda^{-1}$  then $s_W^2(j) \sim 1$ because
$X\star\psi_j(t)$ converges in distribution to a Gaussian white noise of variance $\lambda 2^j$ as $2^j$ goes to $\infty$ \cite{bruna2015intermittent}, and hence
\[
\E\{|X\star\psi_j(t)|\}^2 \sim  \lambda\, 2^{j}~~\mbox{and}~~\E\{|X\star\psi_j(t)|^2\} \sim \lambda\, 2^j~.
\]
The non-self-similarity of Poisson process is also revealed by the large variations of
$C_S(j,a,b)$ along $j$, which appears as large error bars
in Figure \ref{FigEnvelopeCorrelations}.
\paragraph{Multifractal Random Walk (MRW)} It is a non-Gaussian 
self-similar process, whose increments have scaling exponents $\zeta_q$ in (\ref{marginalSelfSimilaritydX})
that are quadratic in $q$. Increments are computed as a product of the increment $\delta_j B$ of a Brownian motion with a log-normal process 
\[
\delta_{j} X(t) = \delta_{j}B(t) \, e^{\Omega_{j}(t)}.
\]
The Gaussian process $\Omega_{j}(t)$ has an auto-correlation with a slow logarithmic decay: 
\[
\Cov\{\Omega_{j}(2^j t), \Omega_{j}(2^j t')\}=\lambda^2 \ln \theta(|t-t'|),
\]
where $\theta$ decreases linearly and is specified in \cite{MRW01}. Since $\Omega_{j}$ is highly correlated in time, it creates wavelet modulus that are also highly correlated in time with bursts of activity. 
The parameter $\lambda$ governs the intensity of this intermittency. 
If $\lambda=0$ then the multifractal random walk is a Brownian motion. 
For MRW one can prove \cite{MRW01} that $\zeta_s = \lambda^2$,
so Table \ref{TableMarginal12Moments} recovers the value $\lambda^2= \zeta_s = 0.016$. 

The scale invariant scattering cross-spectrum
$\overline C_S(a,b)$ is the power spectrum of the modulus auto-correlation at two scales shifted by $a$, where $b$ is a log-frequency index.
Long range correlation of wavelet modulus appears in 
Figure \ref{FigEnvelopeCorrelations}, which shows that $\overline C_S(a,b)$ 
has a power-law decay for each $a$.

A skewed multifractal random walk incorporates a time-asymmetry by imposing
that increments in the past are correlated with the future volatility \cite{pochart2002skewed}, in order to reflect the so-called leverage effect \cite{bekaert2000asymmetric,bouchaud2001leverage}. 
This volatility is defined in finance as the mean square average of 
increments over a fixed period of time. It amounts to replacing the Gaussian process $\Omega_{j}$ by $\Omega_{j} - k \star \delta_{j} B$ where $k(t)$ is a strictly causal power-law convolution kernel. The faster $K$ decreases the shorter the scale of asymmetry. Figure \ref{FigPhasedEnvelopeCorrelations} shows $\overline C_{W|W|}(a)$ for skewed MRW. 
As expected from Proposition \ref{PropTimeReversal1},
this time-asymmetry is revealed by the non-zero phase
of $\overline C_{W|W|}$, which implies that its imaginary part is also non-zero.

\paragraph{Hawkes process} It is a non-homogeneous, causal 
self-excited point process \cite{bacry2014hawkes,bacry2015hawkes}, where
each jump is positive or negative with probability $1/2$. 
The jump intensity $\lambda_t$ depends on the average of the past-jumps with power-law decaying kernel $h$
\[
\lambda_t = \lambda_{\infty} + h \star |d N|_t ,
\]
where $d N_s$ is the signed jump measure and $h \star |d N|_t := \int_{-\infty}^{t}h(t-s)|{\rm d}N_s|$. 
A linear feedback term $l\star dN_t$ can be added to account for correlation between past signed increments and future volatility.

A quadratic Hawkes model \cite{blanc2017quadratic} introduces time-asymmetric modulus dependencies with a causal quadratic feedback of kernel $k(t)$
\begin{equation} 
\nonumber 
\lambda_t = \lambda_{\infty} + h \star |d N|_t + l \star dN_t + |k\star dN_t|^2 .
\end{equation}
If $\int |h(t)| \,dt < 1$ then a quadratic Hawkes is stationary, and if 
$\int (|h(t)| + |k(t)|^2)\,dt < 1$ the mean intensity $\overline{\lambda}$ is finite \cite{Aubrun2022}. 
In numerical simulation, as in \cite{blanc2017quadratic} we set $h(t) \sim |t|^{-1.2}$, $g(t) \sim e^{-0.01 |t|}$
and $k(t) \sim e^{-0.03|t|}$ with $\int (|h(t)| + |k(t)|^2)\,dt = 0.9$.
As expected from Proposition \ref{PropTimeReversal2}, Figure \ref{FigEnvelopeCorrelations}
shows that for this quadratic Hawkes, $\overline C_S(a,b)$ has a non-zero phase 
which reveals its modulus time-asymmetry \cite{zumbach2009time}.

\subsection{Analysis of multi-scale Time-Series}
\label{SubSecAnalysisMultiscaleTimeSeries}

Brownian motions, multifractal random walks and Hawkes processes are used as models of multi-scale time-series, 
particularly in finance and turbulence \cite{mandelbrot1968fractional,mandelbrot1997multifractal,bacry2001modelling,mordant2002long,bacry2014hawkes}. The next two paragraphs analyze the scattering spectra of financial and turbulent time-series. Expected values are computed
with time average estimators, which introduce an estimation error.

\subsubsection{Finance}
\label{SubSecFinance}

Finding stochastic models of financial time-series is important to compute the price of financial instruments which mitigate financial risks, such as options. 
A Brownian motion is a simple first order model, on which the Black and Scholes option pricing formula is based. However, many studies have shown strong deviations to Gaussianity, including the existence of
bursts of activity and crises. Multifractal random walks, Hawkes processes and rough volatility models are among the most popular models used
to capture non-Gaussian properties of financial markets \cite{mandelbrot1963variation,bacry2001modelling,chicheportiche2014fine,roughvol, bacry2015hawkes, blanc2017quadratic}. 
We consider the American stock index S\&P log-prices sampled over 5 minutes from January 2000 to October 2018. A standard preprocessing is performed and is described in \ref{AppendixFinancialDataProcessing}. The resulting signal has $N = 7.5 \, 10^5$ samples. 

Figure \ref{FigMarginalMoments} and Table \ref{TableMarginal12Moments}
show that S\&P has a wavelet spectrum decay exponent $\zeta_2 = 1$, which is the same as a Brownian motion.
However, its sparsity factor $c_s = 0.61 \neq \pi/4\approx 0.79$ and exponent $\zeta_s = 0.016 \neq 0$, which shows that this time-series is clearly
not Gaussian. These values are matched by the MRW model, and by a Hawkes process with calibrated parameters.

The S\&P scale invariant phase-modulus cross-spectrum $\overline C_{W|W|}$ in Figure \ref{FigPhasedEnvelopeCorrelations} is 
similar to a skewed multifractal random walk (SMRW), 
with a strong time-asymmetry shown by a non-zero phase, related to the leverage effect \cite{bekaert2000asymmetric,bouchaud2001leverage}.
The amplitude of $|\overline C_S (a,b)|$ in Figure \ref{FigEnvelopeCorrelations} is similar for the S\&P and
the MRW. However, the phase of $\overline C_S (a,b)$ is non-zero for the S\&P,
which proves that wavelet modulus are also asymmetric in time. This is not well captured by the SMRW model. Such an effect is referred to as the Zumbach effect in finance, which can  be explained 
from causal agent reactions to past trends \cite{zumbach2009time,chicheportiche2014fine, blanc2017quadratic}. Such a modulus time-asymmetry appears in the quadratic Hawkes model, 
but $|\overline C_S (a,b)|$ has different variations along the $a$ direction for S\&P and for the Hawkes model, presumably because of our choice of an exponentially decaying kernel $k(t)$ (which was calibrated on intraday data only).  
This analysis of the scattering spectra shows that none of these mathematical models fully capture all statistical properties of the S\&P time-series.

Financial signals are often believed to be self-similar. We can estimate whether the S\&P satisfies the wide-sense
self-similarity properties of Theorem \ref{main-theo}.
The wavelet spectrum and sparsity factors in 
Figure \ref{FigMarginalMoments} do indeed have a power law decay. Wide-sense self-similarity also imposes that
$C_{W|W|}(0;j,a)$ and $C_S(j,a,b)$ do not depend upon the scale $j$.
Figures \ref{FigPhasedEnvelopeCorrelations} and \ref{FigEnvelopeCorrelations} display their mean-square
variations along $j$ as error bars. They are of the same order as
the estimation variance of each coefficients. We thus conclude that S\&P time-series has no significant deviation from wide-sense self-similarity.

\subsubsection{Turbulence}
\label{SubSecTurbulence}

Kolmogorov introduced in 1941 a self-similar Gaussian model of turbulence 
\cite{kolmogorov1941dissipation,kolmogorov1941degeneration,kolmogorov1941local}, with an asymptotic analysis of Navier-Stokes equations at high Reynolds numbers. This analysis predicts that the projection of the velocity field on a line is a stationary process whose power spectrum has a power-law decay with $\zeta_2 = 2/3$.
However, turbulence time-series are highly non-Gaussian, and Kolmogorov's initial theory was then refined to take into account intermittency and the presence of vortices \cite{kolmogorov_1962,frisch1991global}.

We study the scattering spectra of experimental velocity measurements along a single direction, measured over a
turbulent  gaseous helium jet at low temperature, with a high Reynolds number equal to 929 \cite{chanal2000intermittency}.
This time-series has $N = 3.5\, 10^7$ samples and is thus much larger than the S\&P time-series,
providing more accurate estimators of the scattering spectra.
The non-zero phase of $\overline C_{W|W|}(a)$ and $\overline C_S(a,b)$
in Figures \ref{FigPhasedEnvelopeCorrelations}  and \ref{FigEnvelopeCorrelations}
shows a time-asymmetry, which results from the directionality of the jet propulsion. The quadratic Hawkes provides the best model of $\overline C_S(a,b)$ but it fails to accurately represent $\overline C_{W|W|}(a)$.

It thus appears that none of the existing mathematical model provides accurate models of this turbulent time-series.

Turbulence time-series may be self-similar on a frequency range limited by the Reynolds number at low frequencies and by the fluid viscosity at high frequencies.
The wavelet power spectrum and sparsity factors in Figure \ref{FigMarginalMoments} are indeed
self-similar up to the finest scale $j=2$, which is the diffusion scale created by the fluid viscosity.
The self-similarity error bars
in Figure \ref{FigPhasedEnvelopeCorrelations} and Figure \ref{FigEnvelopeCorrelations}
are of the same order as for the S\&P. However, their amplitude is statistically significant in this case because
this time-series is 50 times larger than the S\&P time-series and the estimation variance is thus much smaller. It shows that this turbulent time-series has significant deviations from wide-sense self-similarity.

\section{Maximum Entropy scattering spectra Models}
\label{SecSynthesisByEntropyMax}

Brownian motions, multifractal random walks and Hawkes models are defined by
a fixed number of parameters which does not depend upon their size $N$. 
They can be calibrated on data, but we saw
they are typically too restrictive to capture all important properties of 
multi-scale time-series. On the other hand, neural network models \cite{oord2016wavenet,eckerli2021generative} have a considerable flexibility but
they can not be trained from a single time-series because the number of parameters
is much larger than $N$. 

This section introduces maximum entropy models computed from the scattering
spectra energy vector $\Phi(x)$ of dimension smaller than $\log^3_2 N$ and hence
much smaller than $N$.
This energy vector is specified in the next section. It provides consistent estimators
of the scattering spectra for random processes having a finite integral scale. 
We study the approximation of multi-scale time-series from such maximum entropy models.
Section \ref{SubSecGenerationFromScatteringModels} shows that scattering spectra models are sufficiently flexible to approximate a wide range of mathematical processes, as well as real financial and turbulence data.

\subsection{Scattering Spectra Energy Vector}
\label{SubSecDiagonalScatteringCv}

We define maximum entropy models of the form
$p_\theta (x) = Z^{-1}_\theta \, e^{-\theta^T\, \Phi(x)}$ for a certain $\theta\in\R^M$, where
the energy vector $\Phi(x)$ computes the scattering spectra
estimation for any  time-series $x$ of dimension $N$.

The scattering spectra vector $\Phi(x)$ 
is normalized by wavelet spectrum coefficients, which
are constants estimated from a realization $\widetilde x$ 
of the random process $X$ that we want to model
\[
\widetilde \sigma_W^2 (j) = \Av{t < N}\big( |\widetilde{x} \star\psi_j(t)|^2 \big) .
\]
If $X(t)$ has a finite integrable scale (see section \ref{subsec:power-spectrum}),
the estimator $\widetilde \sigma_W$ of $\sigma_W$ is consistent as $N$ goes to $\infty$.

Low-frequencies are captured by the low-pass filter $\varphi_J = \psi_{J+1}$ defined in (\ref{lowpass}).
The scattering spectra energy 
is defined by four families of coefficients
\begin{equation} \nonumber
\Phi(x) = \big(\Phi_1(x), \Phi_2(x), \Phi_3(x), \Phi_4(x)\big) .
\end{equation}
The first family provides $J$ order $1$ moment estimators squared, 
corresponding to wavelet sparsity coefficients  (\ref{sparsityCoeff})
\begin{equation}
\label{waveletSpectrum}
\Phi_1(x) = \frac{\big(\Av{t < N} |x \star \psi_j(t)|\big)^2} {\widetilde \sigma_W^2(j)} .
\end{equation}
The $J+1$ normalized second order wavelet spectrum associated to $x$ are computed by
\begin{equation}
\label{sparsityFactor}
\Phi_2(x) = \frac{\Av{t < N} |x \star \psi_j(t)|^2}{\widetilde \sigma_W ^2(j)} .
\end{equation}
There are $J(J+1)/2$ wavelet phase-modulus cross-spectrum coefficients
\begin{equation}
\label{phaseEnvelopeCSpectrum}
\Phi_3(x) = \frac {\Av{t < N} \big( x \star \psi_j(t)\,|x \star \psi_{j-a}(t)| \big)} {\widetilde \sigma_W (j)\, \widetilde \sigma_W (j-a)} .
\end{equation}
Finally, it includes less than $J(J+1)(J+2)/6$ scattering cross-spectrum 
\begin{equation}
\label{scatteringCSpectrum}
\Phi_4(x) = \frac {\Av{t < N}\big(|x\star\psi_{j}|\star\psi_{j-b} (t)\, |x\star\psi_{j-a}|\star\psi_{j-b}^*(t)\big)} {\widetilde \sigma_W (j)\, \widetilde \sigma_W (j-a)} .
\end{equation}
The dimension of $\Phi(x)$ is smaller than $J^3\leq\log^3_2 N$ as soon as $N\geq 8$, this is much smaller than the dimension $N$ of $x$.

\subsection{Model Validation with Test Moments}
\label{SubSecModelValidation}

\begin{figure}
\centering
\includegraphics[width=\linewidth]{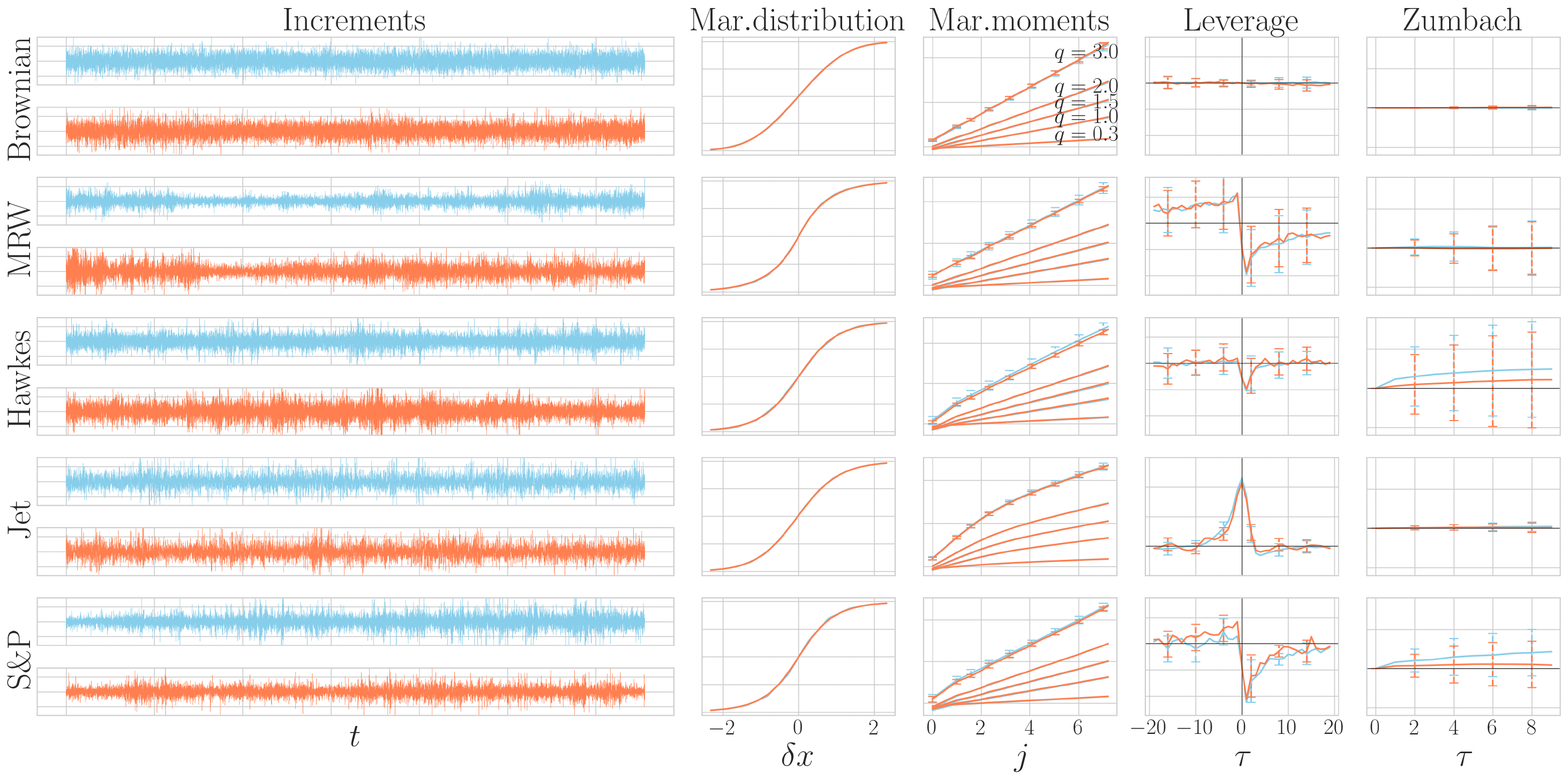}
\caption{Synthesis results. (Left) Increments $\delta_0X(t)$ of original (blue) and generated (orange) time-series. (Right) Using the same color code: cumulative marginal distribution function $F(\delta x)$, log of marginal moments 
${\rm Ave}_{t<N}\,\big(|\delta_j X (t)|^q\big)$ 
as a function of $j$, leverage $\mathcal{L}_j(\tau)$ and Zumbach integral, for the Brownian, MRW (skewed), Hawkes (quadratic), turbulent jet and S\&P. 
Leverage and Zumbach are shown for a specific $j$.
Error bars show the standard deviation for test moments of order greater than 3. Though it is based on order 1 and order 2 moments, our model is able to capture higher order moments used to reveal scaling properties as well as time-asymmetries.
}
\label{FigGeneration}
\end{figure}
This section evaluates numerically the precision of maximum entropy models defined 
from the scattering spectra energy $\Phi(x)$, to approximate the probability distributions of real mathematical processes and real data. The maximum entropy model $p_\theta (x) = Z^{-1}_\theta \, e^{-\theta^T\, \Phi(x)}$ is sampled with a standard microcanonical algorithmic approach reviewed in \ref{app:micro-set}. It is computed with a gradient descent 
which avoids estimating the macrocanoncial parameters $\theta$. 
The choice of the maximum wavelet scale $2^J$ amounts to defining an integrable
scale equal to $2^J$ and hence nearly independent coefficients at distances larger than $2^J$. Indeed, the scattering spectra model then does not impose any constraint on coefficients whose distance are much larger than $2^J$ so the entropy maximisation yields
a random process whose coefficients are nearly independent at such distances.

Assessing the precision of a model from a dataset of samples drawn from an unknown distribution is an ill-defined problem.
A maximum entropy model constrains the values of moments $\E\{\Phi(X)\}$.
One may
find errors by comparing moments which are not explicitly constrained. Such \textit{test moments} are 
estimated on 
the original time-series and on time-series generated by the model itself. 
Following \cite{leonarduzzi2019maximum}, we describe
test moments used in the finance literature to identify important differences relatively to Brownian motions.
Figure \ref{FigGeneration} compares moment values obtained from a scattering spectra model with the original time-series. The following test moments are computed on increments $\delta_j X(t) = X(t) - X(t- 2^j)$ which are stationary. While our model considers dyadic scales $j\in\Z$, we consider all scales $2^j\in\R^+$ for test moments.

{\it Cumulative distribution} $F(\delta x)$ of increments $\delta_0 X(t) = X(t) - X(t-1)$ at the finest scale. 

{\it Marginal high order moments} described in (\ref{variationsOrderq}) are estimated by time average $\Av{t < N}\big(|\delta_j X (t)|^q\big)$ at all scales $2^j$, for $0.3 \leq q \leq 3$. The multi-scale properties of these moments
is reviewed in Section \ref{SubSecHighOrderMomentsandSelfSimilarity}.

{\it Leverage moments} measure asymmetric dependencies between past 
and future increments in finance \cite{bekaert2000asymmetric,bouchaud2001leverage}. A leverage correlation 
for a time shift $\tau$ is an order 3 moment, at any scale $2^j$:
\[
\mathcal{L}_j(\tau) = \Av{t < N}\big(\delta_j X(t-\tau)|\, \delta_j X(t)|^2\big). 
\]
If $X$ has a time-reversible distribution then $\mathcal{L}_j(\tau) =-\mathcal{L}_j(-\tau)$.
Results are shown in Figure \ref{FigGeneration} at an intermediate scale $2^j=159$ that corresponds to the day in the case of S\&P. The same scale is taken for all processes except for the Jet for which we take $2^j=1$.

{\it Zumbach moments} evaluate the time-asymmetry of the volatility
in finance \cite{zumbach2009time,chicheportiche2014fine}. The volatility is
the energy of increments over a time period of size $2^j$
\[
\Sigma_j^2 (t) = \Av{t-2^j\leq u<t} |\delta_0 X(u)|^2 . 
\]

A Zumbach moment for a time shift $\tau$
is an order $4$ moment at a scale $2^j$
\[
\mathcal{Z}_j(\tau) = \Av{t < N}\big(|\delta_j X(t-\tau)|^2 \,\Sigma_j^2 (t)\big) .
\]
If $X$ is time-reversible then $\mathcal{Z}_j(-\tau)=\mathcal{Z}_j(\tau)$. To evaluate the time-asymmetry, Figure \ref{FigGeneration} shows
$\int_0^t(\mathcal{Z}_j(s) - \mathcal{Z}_j(-s))\, \mathrm{d}s$ as a function of $t$ for a scale $2^j=159$, that corresponds to the day for S\&P. This asymmetry coefficient on an order $4$ moment is typically estimated with a large variance as we shall see.

\subsection{Generation from Scattering Spectra Models}
\label{SubSecGenerationFromScatteringModels}

Figure \ref{FigGeneration} gives results of scattering spectra models computed 
from a single realization of a Brownian motion, a skewed MRW, a quadratic Hawkes process, a turbulence jet and the S\&P financial signal. It displays realizations generated by these models and compares test moments. For financial and turbulence data, the syntheses recover
signals of size $N=7.10^5$ with models computed over $J=11$ scales. 
The 
resulting scattering spectra model has $375$ parameters. 
Figure \ref{FigGeneration} shows that all test moments of order $2$ or below are perfectly reproduced by
scattering spectra models, for Brownian motion, MRW and Hawkes as well as for the turbulence jet and S\&P financial data. Marginal moments of order $1/3 \leq q \leq 3$ are captured by our model on all processes, which is in accordance with the well reproduced cdf.

For test moments of order $3$ or $4$, including the leverage and Zumbach effects that capture time-asymmetries, we represent the variance of estimators with an error bar, which is quite large for the Zumbach effect. 
Leverage is well captured for MRW, Hawkes, Jet, and remains within the estimation error bar for S\&P. Zumbach integral estimations have a much larger variance. The main information is in the sign of this integral, when significant. This test moment is again reproduced on both Hawkes and S\&P within the estimation error. Its high variance clearly shows the importance of using low order moments, even for time-asymmetries. scattering spectra models reveal such non-Gaussian properties with a modulus and moments of order $1$ and $2$. 
In the case of the S\&P, we believe that any remaining discrepancy for the Zumbach effect comes from our somewhat naive treatment of closing periods during the night.

\section{Conclusion}

We introduced the scattering spectra which give an interpretable low-dimensional representation of processes having stationary increments. It captures their power spectrum, multi-scale sparsity, and the dependencies of wavelet coefficients phase and modulus across scales.
Wide-sense self-similar signals have scattering spectra which are invariant to scale shifts, and thus
define a representation of even lower dimension. 

For a time series of size $N$, this scattering spectra is at most of dimension $\log_2^3 N$. We showed numerically that it reveals potential non-Gaussianity and self-similar properties. This was demonstrated on mathematical multi-scale models such as fractional Brownian motions, multifractal random walks and Hawkes processes, but also on real time-series in finance and turbulence. Maximum entropy scattering spectra models capture essential multi-scale dependency properties and can be efficiently sampled with a microcanonical approach.

Scattering spectra models are related to generative convolutional neural networks based on covariance matrices \cite{gatys2015texture}. Similarly to a one-hidden layer
convolutional neural network, it computes a cascade of two convolutions and a pointwise non-linearity. The network filters are wavelets which are not learned. It provides a much lower dimensional representation of random processes than usual deep convolutional neural networks, and it is furthermore interpretable. However, it only applies to signals which are stationary or have stationary increments.


\newpage

\appendix

\section{Wavelet Transform Properties}
\label{AppendixWaveletProperties}

We impose that the wavelet $\psi$ satisfies the following energy conservation law called Littlewood-Paley equality
\begin{equation}
\label{littlewood}
\forall \om > 0~~,~~ \sum_{j=-\infty}^{+\infty} |\widehat {\psi} (2^j \om)|^2 = 1 .
\end{equation}
A Battle-Lemarié wavelet \cite{battle1987block, lemarie1988ondelettes} is an example of such wavelet. 
The wavelet transform is computed up to a largest scale $2^J$ which is smaller than the signal size $N$.
The signal lower frequencies in $[-2^{-J} \pi, 2^{-J} \pi]$ are captured by a low-pass filter $\varphi_J(t)$ whose Fourier transform is
\begin{equation}
\label{lowpass}
\widehat{\varphi}_J (\om) = {\Big(\sum_{j=J+1}^{+\infty} |\widehat{\psi} (2^j \om)|^2\Big)^{1/2}} . 
\end{equation}
One can verify that it has a unit integral $\int \varphi_J(t)\, {\rm d}t = 1$. 
To simplify notations, we write this low-pass filter as a last scale wavelet
$\psi_{J+1} = \varphi_J$, and $W x(t,J+1) = x \star \psi_{J+1}(t)$. By applying the Parseval formula,
we derive from (\ref{littlewood}) that for all $x$ with $\|x\|^2 = \int |x(t)|^2\,dt < \infty$
\begin{equation}
\nonumber
\|W x \|^2 = \sum_{j=- \infty}^{J+1}\|x \star \psi_j \|^2 = \|x\|^2 .
\end{equation}
The wavelet transform $W$ preserves the norm and is therefore invertible, with a stable inverse.

Properties of signal increments are carried over to wavelet coefficients by observing that wavelet coefficients are obtained by filtering signal increments
$\delta_j X(t) = X(t) - X(t - 2^j)$ with a dilated integrable filter:
\begin{equation}
\label{dXvsWx}
X \star \psi_j (t) = \delta_j X \star \theta_j (t)~~
\mbox{where}~~ \theta_j (t) = 2^{-j} \theta(2^{-j} t),
\end{equation}
where filter $\theta$ is obtained from $\psi$ through
$\widehat{\theta} (\om) = \widehat{\psi}(\om)\, / \, (1 - e^{-i \om})$. 
This is because $1-e^{i2^j\omega}$ is the Fourier transform of $\delta(t) - \delta(t-2^j)$, the filter that creates increments.
From (\ref{dXvsWx}) we get that if $\delta_j X(t)$ is stationary then $X \star \psi_j (t)$ is also stationary.

\section{Proof that strong distribution self-similarity implies weak moment self-similarity.}
\label{AppendixDistributionImpliesMarginal}

Let $X$ be a stationary process that is self-similar according to (\ref{strongSelfSimilaritydX}). For $q\in\R$, marginal moments are written $S(q,j) = \E\{|\delta_jX(t)|^q\}$. They do not depend upon $t$. For all $\ell\geq0,j\leq J$, self-similarity implies that marginal distributions are equal: $\delta_jX(2^\ell t) \overset{d}{=} A_\ell\,\delta_{j-\ell}X(t)$. On order $q$ moments, since $A_\ell$ is independent from $X$ this yields
\[
S(q,j) = \E\{A_\ell^q\} S(q,j-\ell).
\]
Since the factors $(A_\ell)_\ell$ are log-infinitely divisible, for all $\ell_1,\ell_2\geq0$ $\E\{A^q_{\ell_1+\ell_2}\}=\E\{A^q_{\ell_1}\}\E\{A^q_{\ell_2}\}$. This implies that $\log \E\{A^q_{\ell}\}$ is linear in $\ell$ which means there exists $\zeta_q$ such that $\E\{A^q_{\ell}\}=2^{j\zeta_q}$. By defining 
$\widetilde S(q,j)=2^{-j\zeta_q}S(q,j)$, we obtain $\widetilde S(q,j)=\widetilde S(q,j-\ell)$ for all $j\leq J,\ell\geq0$, which implies that $\widetilde S(q,j)$ is equal to a constant $\widetilde c_q$ which 
does not depend on $j$
\begin{equation}
\label{dXpowerlaw}
S(q,j)=\E\{|\delta_j X(t)|^q\} = \widetilde c_q 2^{j\zeta_q}.
\end{equation}

Let us now establish the same property for wavelet coefficients. According to the self-similarity property (\ref{strongSelfSimilaritydX}), wavelet coefficients satisfy (\ref{strongSelfSimilarityWX}). Indeed, one has:
\begin{align*}
X\star\psi_j(2^\ell t) & = \delta_j X \star\theta_j(2^\ell t) ~~ \text{thanks to} ~~ (\ref{dXvsWx})
\\
& = \delta_j X (2^\ell \cdot) \star \theta_{j-\ell} (t) ~~ \text{as $\theta_j$ are dilated filters}
\\
& \overset{d}{=} A_\ell \, \delta_{j-\ell} X \star \theta_{j-\ell} (t) ~~ \text{by self-similarity} ~~ (\ref{strongSelfSimilaritydX})
\\
& = A_\ell \, X\star\psi_{j-\ell}(t) ~~ \text{thanks to} ~~ (\ref{dXvsWx})
\end{align*}
Increments $\delta_j X(t)$ are a special case where $\psi_j=\delta(t) - \delta(t-2^j)$. For general wavelets $\psi_j$ the same proof than for increments holds,
there exists $c_q$ such that for the same $\zeta_q$:
\[
\E\{|X\star\psi_j(t)|^q\} = c_q 2^{j\zeta_q} .
\]

\section{Proof that strong distribution self-similarity implies wide-sense self-similarity}
\label{AppendixProofTheorem}

Let $X$ be a process with stationary increments that is self-similar and thus satisfies (\ref{strongSelfSimilarityWX}). For $q=1$ and $q=2$, \ref{AppendixDistributionImpliesMarginal} proves that $\E\{|X \star \psi_j (t)|\}  =  c_1\, 2^{j \zeta_1}$ and
\begin{equation}
\label{selfSimilarityWaveletSpectrum}
\sigma_W^2(j) = \E\{|X \star \psi_j (t)|^2\}  =  c_2\, 2^{j \zeta_2},
\end{equation}
which proves (\ref{DefOrder1}) and (\ref{DefOrder2}). 
\\
The equality in distribution (\ref{strongSelfSimilarityWX}) implies that
for $\tau,j,a$ fixed we have
\begin{align*}
\Big(X\star\psi_{j}(t), \,& X\star\psi_{j-a}(t-2^{j}\tau)\Big)
\\
\overset{d}{=}
&
A_\ell\Big(X\star\psi_{j-\ell}(2^{-\ell}t), \, X\star\psi_{j-a}(2^{-\ell}t-2^{j-\ell}\tau)\Big)
\end{align*}
Applying $\rho$ and taking expected value gives
\begin{align*}
& \E\{\rho WX(t,j)\, \rho WX(t-2^j\tau,j-a) \}
\\
& = 2^{\ell\zeta_2} \E\{\rho WX(t,j-\ell) \,
\rho WX(t-2^{j-\ell}\tau,j-a-\ell)\},
\end{align*}
because of stationarity and $\E\{A_\ell^2\}=2^{\ell\zeta_2}$. 
Normalized correlations $C_{\rho W}(\tau;j,a)$ are obtained by dividing by $\sigma_W(j)\sigma_W(j-a)$.
It results from (\ref{selfSimilarityWaveletSpectrum}) that 
\[
2^{\ell\zeta_2} \sigma_W(j)^{-1}\sigma_W(j-a)^{-1}=\sigma_W(j-\ell)^{-1}\sigma_W(j-a-\ell)^{-1}
\]
and hence
\[
C_{\rho W}(\tau;j,a) = C_{\rho W}(\tau,j-\ell,a).
\]
Taking $j=\ell$ proves (\ref{DefCross}). 

\section{Proof of proposition \ref{PropTimeReversal2} and theorem \ref{sndTheorem}}
\label{AppendixProofOfPropositionTimeReversalAndTheorem}

Let $X$ be a Gaussian process with stationary increments and assume that $\widehat\psi_j \widehat\psi_{j-a} = 0$. Then for any $\tau$, $X\star\psi_j(t)$ and $X\star\psi_{j-a}(t-\tau)$ are decorrelated because their power spectra do not overlap. Since $X$ is Gaussian these are also Gaussian. It implies that the processes $X\star\psi_j(t)$ and $X\star\psi_{j-a}(t)$ are independent. In particular,  $|X\star\psi_j|\star\psi_{j-b}(t)$ and $|X\star\psi_{j-a}|\star\psi_{j-b}(t)$ are independent. Their correlation is thus zero and $C_S(j,a,b)=0$.

Let $X$ be a time-reversible process with stationary increments.
Thanks to proposition \ref{PropTimeReversal1} we know that $C_{\rho W}$ has the Hermitian symmetry:
$C_{\rho W}(\tau;j,a)=C_{\rho W}(-\tau;j,a)^*$. 
Let us write $C_{|W|}$ the subblock of this matrix composed of the normalized modulus auto-correlation
\[
C_{|W|}(\tau;j,a) =
\frac{
\E\{|X\star\psi_j(t)| \, |X\star\psi_{j-a}(t-2^j\tau)|\}
}
{\sigma_W(j) \sigma_W(j-a)}
\]
We derive from (\ref{scatProjection}) that the scattering cross-spectrum $C_S$ satisfies:
\begin{equation}
    \label{cmwTocs}
    C_S(j,a,b) =  \int_{\tau} C_{|W|}(\cdot \, ;j,a) \star \psi_{-b}(\tau) \ \psi_{-b}(\tau)^* {\rm d} \tau.
\end{equation}
With $Rx(t)=x(-t)$, the Hermitian symmetry of $C_{|W|}$ implies that
\begin{align*}
    C_S(j,a,b) & = \int_{\tau} RC_{|W|}(\cdot \, ;j,a)^* \star \psi_{-b}(\tau) \ \psi_{-b}(\tau)^* {\rm d} \tau
    \\
    & = \int_{\tau} C_{|W|}(\cdot \, ;j,a)^* \star R\psi_{-b}(-\tau) \ \psi_{-b}(\tau)^* {\rm d} \tau
    \\
    & = \int_{\tau} C_{|W|}(\cdot \, ;j,a)^* \star \psi^*_{-b}(-\tau) \ \psi_{-b}(\tau)^* {\rm d} \tau
    \\
    & = \int_{\tau} C_{|W|}(\cdot \, ;j,a)^* \star \psi^*_{-b}(-\tau) \ \psi_{-b}(-\tau) {\rm d} \tau
    \\
    & = C_S^*(j,a,b)
\end{align*}
because $R\psi_{-b} = \psi_{-b}^*$. It proves that ${\rm Im} \, C_S(j, a, b) = 0$ which proves proposition \ref{PropTimeReversal2}.

Under scale invariance (\ref{DefCross}) $C_{\rho W}(\tau; j,a) = C_{\rho W}(\tau; 0,a)$. As a subblock, one has also $C_{|W|}(\tau; j,a) = C_{|W|}(\tau; 0,a)$. In particular, the equation (\ref{cmwTocs}) that expresses $C_S$ from $C_{|W|}$ implies $C_S(j,a,b)=C_S(0,a,b)$ which proves theorem \ref{sndTheorem}.

\section{Financial data preprocessing}
\label{AppendixFinancialDataProcessing}

We use a standard preprocessing on S\&P data which accounts for missing values, overnight period, intraday seasonality and tick effect. It is performed on 5min increments of S\&P from January 3rd 2000 to October 10th 2018 which represents 751\,116 values. 

Missing values, 17\,956 5min increments, are replaced by independent Gaussian values with zero mean and standard deviation observed at this time of the day.

Intraday seasonality is the fact that the volatility is larger at certain typical hours of the day: it is a non-stationary effect. It is removed by dividing 5min increments by the average volatility profile over all days.

The overnight period corresponds to the first bin at the beginning of each day. The corresponding increments are generally larger than other 5min increments. That again creates a non-stationarity effect that can be attenuated by dividing each overnight increment by their average volatility on all days.

Prices of S\&P are present on a grid with certain tick size. Hence, 5min increments are discrete with many values equal to 0 or to plus/minus the tick size. This tick effect is present only for high-frequency increments and hence breaks the scale invariance property at high frequency. To remove it we apply a low-pass filter to $X$ that amounts to a moving average on small windows of 15 minutes.

\section{Microcanonical set and sampling}
\label{app:micro-set}

We briefly review a maximum entropy microcanonical approach to
sample a maximum entropy model defined from a single realization $\widetilde x$ of $X$ \cite{bruna2019multiscale}.
A microcanonical set of width $\epsilon$
takes into account the error between $\Phi(\widetilde{x})$ and $\E\{\Phi(X)\}$ 
\[
\Omega_\epsilon = \{ x\in\R^N \ | \ \| \Phi(x) - \Phi(\widetilde{x})\|_2  \leq \epsilon  \}.
\]
The width $\epsilon$ is adjusted so that each realization of $X$ of size $N$ is in $\Omega_\epsilon$, with high probability (in practice, we choose $\epsilon=10^{-3}\|\Phi(\widetilde x)\|_2$). The probability distribution of $X$ is thus essentially supported in $\Omega_\epsilon$. The width $\epsilon$ goes to zero when $N$ increases if $\Phi(X)$ is a mean-square consistent estimator of $\E\{\Phi(X)\}$ when
$N$ goes to infinity. The convergence of microcanonical models towards macrocanonical models is a large deviation problem studied in statistical physics  in the limit of $N$ going to infinity \cite{lanford1975time}.

A microcanonical model is the maximum entropy distribution supported in $\Omega_\epsilon$. 
Since $\Omega_\epsilon$ is bounded, the microcanonical model has a uniform distribution over $\Omega_\epsilon$.
Sampling the microcanonical model thus amounts to select $x$ in 
$\Omega_\epsilon$, with a uniform probability.
Such a sampling can be calculated by a gradient descent studied in 
\cite{bruna2019multiscale}. It progressively transports a Gaussian white noise, which has a maximum entropy, into a distribution supported in $\Omega_\epsilon$, with a push-forward map calculated with a gradient descent on $\ell(x) = \| \Phi(x) - \Phi(\bar x)\|^2_2$.

The initial point $x_0$ is a realization of a Gaussian white noise. At each iteration $n$, it updates $x_n$ with a gradient descent step
\[
x_{n+1} = x_n - \eta \nabla \ell(x_n).
\]
The gradient descent is implemented with the L-BFGS-B algorithm. It is proved in \cite{bruna2019multiscale} that
it converges to a measure which approximates the maximum entropy measure. 
It has the same 
invariance to time shift
as $\Phi(x)$ similarly to the maximum entropy measure.




\end{document}